\UseRawInputEncoding
\documentclass[12pt,a4paper]{article}
\usepackage{geometry}
\usepackage[round]{natbib}
\usepackage{mathtools}
\usepackage{amsfonts}
\usepackage{amssymb}
\usepackage{microtype} 
\usepackage{graphicx}
\usepackage{setspace}
\usepackage{placeins}
\usepackage{epstopdf}
\usepackage[small,labelfont=bf]{caption}
\usepackage[skip=-0.25\textwidth]{subcaption}
\usepackage{enumitem}
\usepackage{titlesec}
\titleformat{\section}
{\normalfont\fontsize{12}{15}\bfseries}{\thesection .}{0.3em}{}
\titleformat{\subsection}
{\normalfont\fontsize{12}{15}\itshape\bfseries\centering}{\upshape\thesubsection .}{0.3em}{}
\titleformat{\subsubsection}
{\normalfont\fontsize{12}{15}\itshape\bfseries}{\upshape\thesubsubsection .}{0.3em}{}

\usepackage{color}
\usepackage[titletoc,title]{appendix}
\usepackage{xpatch}
\xpretocmd{\eqref}{equation\,}{}{}
\usepackage{authblk}

\title{\textbf{\Large Steering a thermally activated micromotor  with a nearby isothermal wall}}
\author[1]{Antarip Poddar}
\author[1]{Aditya Bandopadhyay\thanks{Email: aditya@mech.iitkgp.ernet.in}}
\author[1]{Suman Chakraborty\thanks{Email: suman@mech.iitkgp.ernet.in}}
\affil[1]{Department of Mechanical Engineering, Indian Institute of Technology Kharagpur, Kharagpur, West Bengal - 721302, India}
\date{}                     
\begin{document}
            \maketitle
\begin{spacing}{1.15}
\begin{abstract}
\noindent
Selective heating of a microparticle surface had been observed to cause its autonomous movement in a fluid medium due to self-generated temperature gradients. In this work, we theoretically investigate the response of such an auto-thermophoretic particle near a planar wall which is held isothermal. We derive an exact solution of the energy equation and employ the Reynolds reciprocal theorem to obtain the translational and rotational swimming velocities in the creeping flow limit. Subsequently, we analyse the trajectories of the micromotor for different  thermo-physical and  configurational parameters. Results show that the micromotor trajectories can be switched either from wall-bound sliding or stationary state to escape from the near-wall zone by suitably choosing the particle and the surrounding fluid pair having selective thermal conductivity contrasts. Further, we discuss the dependency of this swimming-state transition on the launching orientation and the coverage of the metallic cap. Our results reveal that the scenario addressed here holds several exclusive distinguishing features from the otherwise extensively studied self-diffusiophoresis phenomenon near an inert wall, despite obvious analogies in the respective constitutive laws relating the fluxes with the gradients of the concerned forcing parameters. The most contrasting locomotion behaviour here turns out to be the ability of a self-thermophoretic micromotor to migrate towards the wall with large heated cap even if it is initially directed away from the wall. Besides, during the stationary state of swimming, the cold portion on the micromotor surface faces away from the wall, under all conditions. Such unique aspects of locomotion hold the potential of being harnessed in practice towards achieving intricate control over autonomous motion of microparticles in thermally-regulated fluidic environments.
    \end{abstract}
\end{spacing}    
\par\noindent\quad\rule{0.95\textwidth}{0.4pt}

\section{Introduction}
The exclusiveness of artificial mircoswimmers in harnessing energy from the surrounding fluid and converting that to useful mechanical energy for propulsion has made them suitable for a wide range of potential applications that include but are not limited to targeted drug delivery \citep{Wang2012}, disease diagnosis \citep{Chalupniak2015}, environmental remediation \citep{Li2014a}, and photothermal therapy \citep{Choi2018}. These microswimmers often take the form of micron-sized colloidal particles (also known as Janus particles) having asymmetric properties of their two faces. This asymmetric distribution of transport characteristics essentially triggers self-generated chemical, electrical, optical, or thermal gradients \citep{Paxton2006,Howse2007,Lozano2016,Qian2013,Jiang2010}, leading to exclusive propulsion characteristics despite the absence of any external forcing.

Among the different propulsion mechanisms, the chemical decomposition of hydrogen peroxide has been extensively studied in the literature \citep{Paxton2004,Paxton2006,Gibbs2009,Wang2015,Poddar2019}. However, its use in biomedical applications is limited due to the toxicity of the fuel and requirement of continuous supply of the same in the environment. Fuel-free locomotion can, however, be realised by localised heating of a metallic cap engendering a temperature gradient, which causes a directed self-thermophoretic locomotion  \citep{Jiang2010}. The heating can be accomplished by different external stimuli such as illumination of laser beams \citep{Jiang2010,Ilic2016,Qian2013,Bregulla2014} or employing an alternating magnetic field \citep{Baraban2013}. Opto-thermal steering of the micromotor can also be performed by layering the two faces with different light-absorbing materials \citep{Ilic2016}. This coating pattern renders the local temperature gradient bidirectional, and selective lighting can result in more control over the autonomous movement. Recently, a rotating electric field has also been employed to enhance the self-generated motion due to intrinsic thermophoresis \citep{Chen2018}.

Despite the obvious physical distinctions, thermally-driven and solutally-driven transport phenomena are commonly attributed to several conceptual similarities. In fact, it has been established that temperature and solute concentration fields as well as the fluid flow patterns around an autophoretic particle in an unconfined fluid domain bear extreme qualitative similarities, primarily due to analogies in the respective constitutive laws relating the fluxes with the gradients of the forcing parameter, and often a common framework of analysis may be adopted \citep{Golestanian2007} to probe the pertinent implications.  Exclusive geometrical and physical implications of a nearby wall, in addition, may influence the concerned aspects of particle locomotion, to a significant extent. 
 Inspired by the experimental evidence about the capability of a nearby wall in achieving precise navigation of a chemically active particle \citep{Das2015}, a host of studies has been aimed at analysing the motion of a self-diffusiophoretic micromotor near a solute-impermeable plane wall \citep{Mozaffari2016,Uspal2015a,Ibrahim2016,Crowdy2013}. 
 However, taking cues from previously studied problems on passive particle thermophoresis near a wall \citep{Chen1999,Chen2000a,Chen2000}, it can be realised that an isothermal wall is likely to bring in unique artefacts to its thermally-modulated dynamical evolution, which, by no means may be extrapolated trivially from other previously reported observations on diffusiophoretic transport in a wall-bounded flow. This may be attributed to the fact that the physical similitude of solute-impermeability condition for a diffusiophoresis problem leads to an equivalent adiabatic paradigm for a thermophoresis problem, which grossly deviates from an isothermal scenario.
 
  In addition, contrasts in thermal conductivity between the particle and surrounding fluid are likely to result in key dynamical evolutions by virtue of altering the incipient thermo-hydrodynamic characteristics \citep{Jiang2010,Bickel2013}. While the implications of a nearby isothermal wall on the later are presumably very complicated, the aspect of thermal conductivity gradient-driven transport of an active micromotor by itself has turned out to be an unaddressed phenomenon, even in an unbounded flow domain.

 Here, we aim to unveil the physical consequences of a self-thermophoretic microswimmer adjacent to an isothermal plane wall; an aspect of active particle hydrodynamics that has hitherto remained unaddressed. The coupling between the thermal and the hydrodynamic field, mediated by kinematic and kinetic constraints and thermal boundary conditions, is portrayed to be the key in dictating the unique dynamical evolution under this purview. Instead of solving the full velocity profile of the fluid, we make use of the Reynolds reciprocal theorem \citep{Happel1983} to evaluate the phoretic thrust experienced by the particle and employ the force-free conditions in obtaining the translational and rotation velocities of the swimmer. The exact solution of the energy equation is obtained by considering steady-state conditions and neglecting temperature distortion due to fluid advection. The results indicate that the thermal conductivity contrast between the particle and the surrounding fluid serves as a switching mechanism for the micromotor trajectories. With the interplay of the consequent forcing with the wall-induced thermophoresis, we further bring out multiple characteristics of near-wall swimming states that stand apart from the widely studied problem of self-diffusiophoresis in the proximity of an inert wall, and hold the key in opening up several novel applications featuring thermally-activated control of active matter in a wall-bounded flow medium.

\section{Mathematical formulation}
\label{sec:Mathematical}
\begin{figure}[!htb]
    \centering
    \includegraphics[width=0.7\textwidth]{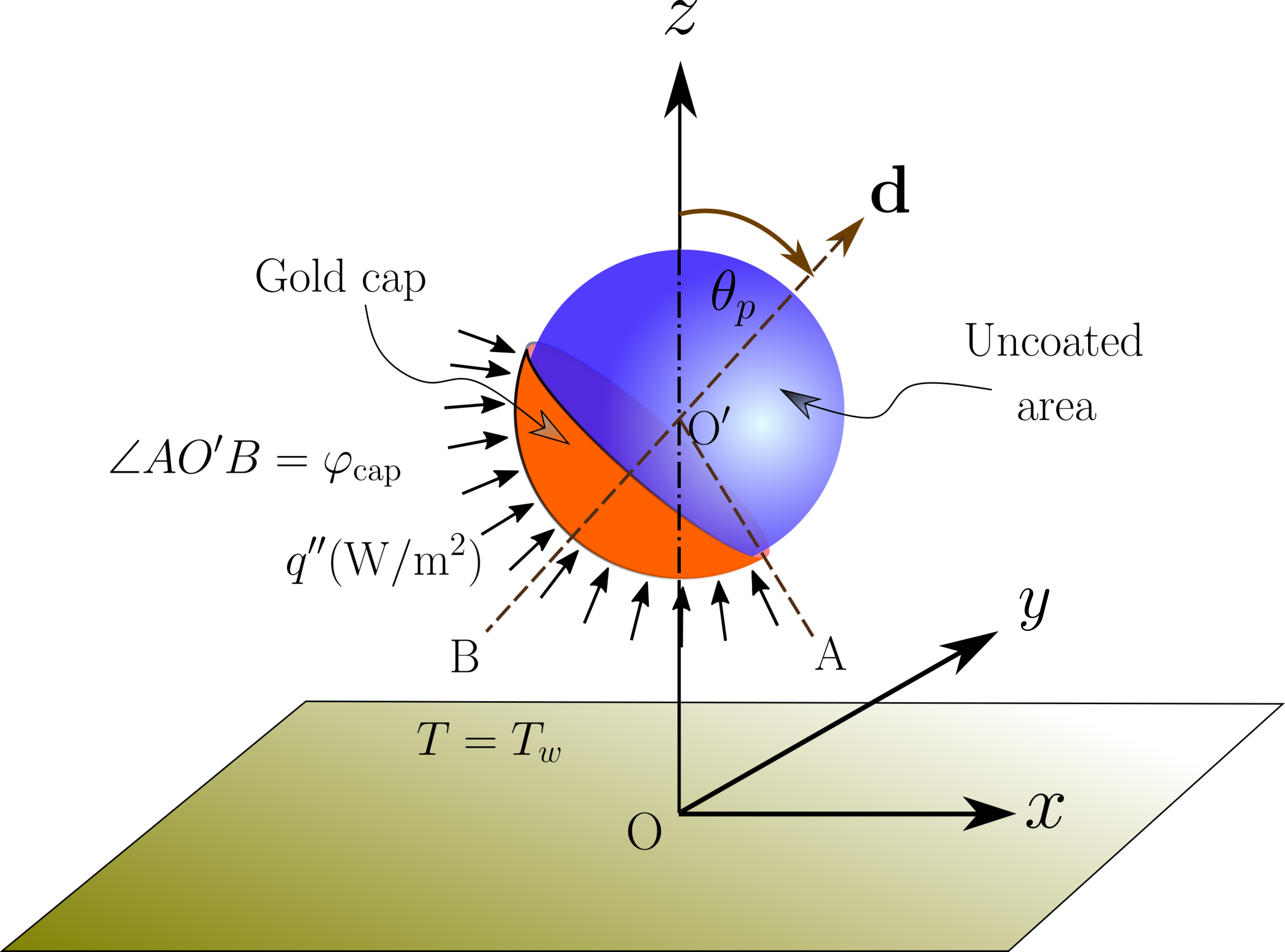}
    \vspace{2ex}
    \caption{Schematic of an auto-thermophoretic microswimmer near a constant temperature plane wall. The orange and blue faces of the particle represent the heat-absorbing metallic cap (hot) and uncoated (cold) areas on the micromotor surface, respectively. The angular orientation of the director vector $ \mathbf{{d}} $ is denoted by the angle $ \theta_{p} $, measured in the clockwise direction from the positive $z$-axis. The coverage area of the metallic cap is represented by the angle $ \varphi_\textrm{cap} $. The centre of the particle is at a distance of $ \widetilde{h} $ from the wall.}    
    \label{fig:schematic_temp}
\end{figure}

We consider a micron-size Janus particle of thermal conductivity $\varkappa_p$ in a fluid of thermal conductivity $\varkappa_f$.
In typical cases, the particle is partially with gold (Au) or titanium-nitride (TiN), which absorbs light of specific frequency \citep{Jiang2010,Ilic2016}. This coating is assumed to be axisymmetric about $ \mathbf{{d}}$, as shown in the figure~\ref{fig:schematic_temp}. Upon irradiation of laser beam, an asymmetric conversion to thermal energy takes place along the coated and uncoated faces of the particle. This triggers a local temperature gradient around the particle, and subsequently, a difference in osmotic pressure is created. Correspondingly, a slip flow is developed along the particle surface, and while viewing from the laboratory frame, the particle is seen to propel itself. In an unbounded domain, the propulsion is directed along the unit vector $ \mathbf{{d}}$. Such phenomenon comprises the basics of an auto-thermophoretic microswimmer \citep{Jiang2010,Bickel2013}, albeit without any interactions with a nearby bounding wall. In the proximity to a constant temperature $(T_w)$ plane wall, this thermophoretic motion is likely to be altered significantly, which is considered to be the focal point of the subsequent analysis.

For theoretical depiction, we adopt the bispherical co-ordinate system $ (\xi, \eta, \phi).$ This is related to the cylindrical system $ (\rho,z,\phi) $ (with origin at the plane wall and the $ z $ axis being directed normal to the wall and passing through the swimmer center) as \citep{Happel1983}
\begin{equation}\label{eq:bisp_def}
\widetilde{\rho}=c\frac{\sin(\eta)}{\cosh(\xi)-\cos(\eta)} \quad \text{and} \quad \widetilde{z}=c\frac{\sinh(\xi)}{\cosh(\xi)-\cos(\eta)}.
\end{equation}
Here $ c $ is a positive scale factor, $ \xi=0 $ denotes the plane wall location, and $ \xi=\xi_0 $ ($>0$) represents  the spherical swimmer surface. The sphere has its center situated  at a height of $ \widetilde{z}=\widetilde{h}=c \, \coth(\xi_0)$, and it has a radius of $a= c/\sinh(\xi_0)$. Thus, the smallest distance between the sphere-surface and plane wall becomes $ \delta=h-1=c \, \coth(\xi_0)/a-1$.
\subsection{Temperature distribution in and around the microswimmer}
\label{ssec:temp_distr}
We neglect the flow-induced distortions in the temperature field from a co-moving reference frame and consider rapid diffusion of thermal energy, leading to a quasi-steady-state behaviour of the temperature distribution \citep{Golestanian2007}. Hence, the energy equations in both the particle $ (p) $ and the surrounding fluid $ (f) $ phases reduce to
\begin{equation}\label{eq:energy_eq}
\nabla^{2}T_{i} = 0,\qquad i=p,f.
\end{equation}
We leverage on facts that the laser-absorbing metallic cap, in many experimental scenarios, is only tens of nanometres in thickness $ (t_{cap}) $, and that the particle radius $(a)$ lies in the range of microns \citep{Jiang2010,Ilic2016}. Moreover, the thermal conductivity of the cap is much larger than  that of either of the particle or the fluid. Thus, the condition $\varkappa_{cap}t_{cap} \gg \varkappa_p a$ is satisfied, and  following the earlier works \citep{Jiang2010,Bickel2013}, it may be judicious to assume the `thin cap limit'. The thermal energy emitting from the laser-absorbing metallic cap causes a jump in the heat flux across the coated portion of the particle, while along the uncoated colder portion, a simple continuity of the heat flux is maintained. This is mathematically described by employing the following boundary condition at the particle surface:
\begin{equation}\label{eq:bc_1a}
\text{at} \;\xi=\xi_0, \qquad -\varkappa_{f}(\nabla T_{f})\cdot\mathbf{{n}} + \varkappa_{p}(\nabla T_{p})\cdot\mathbf{{n}} = \mathcal{Q}(\mathbf{{n}}),
\end{equation}
where $ \mathcal{Q}(\mathbf{{n}}) $ is a piecewise function depicting the local rate of heat absorption per unit area of the particle surface, defined as
\begin{equation}
\mathcal{Q}(\mathbf{{n}})=%
\begin{cases}
0, &\text{if} \; \cos(\pi- \varphi_\text{cap}) \le \mathbf{{d} \cdot {n}} \le 1 \\
q'', &\text{otherwise}.
\end{cases}
\end{equation}
The temperature is held fixed at the nearby flat wall $(T_w)$.
Without loss of generality, we set the wall temperature to zero with respect to a suitably chosen reference temperature, i.e.
\begin{equation}\label{eq:bc_wall}
\text{at} \quad \xi=0,\quad T_w=0.
\end{equation}
Further, the temperature gradient must vanish at large distances from the particle in the domain $ z > 0. $

In terms of the eigenfunctions in the bispherical coordinates, the temperature field in the surrounding fluid medium can be expressed as a solution of the Laplace equation (\ref{eq:energy_eq}) \citep{Jeffery1912,Subramanian2001}, as given below:
\begin{align}
\label{eq:Temp_out_1}
T_{f} = \sqrt{\cosh \xi - \cos\eta} \sum_{n=0}^{\infty} \sum_{m=0}^{\infty}\big[A_{n,m}\sinh(n+1/2)\xi +B_{n,m}\cosh(n+1/2)\xi \big] \nonumber \\
\times P_n^{m}(\cos\eta)\cos(m\phi + \gamma_m).
\end{align}
{We emphasize the fundamental differences of the temperature distribution with that of the concentration profile of a closely related self-diffusiophoresis problem \citep{Mozaffari2016}. First, the boundary condition at $ \xi=0 $ (\eqref{eq:bc_wall})
    gives $B_{n,m}=0.$
    Thus, the outer region temperature profile reduces to
    \begin{equation}
    \label{eq:Temp_out_2}
    T_{f} = \sqrt{\cosh \xi - \cos\eta} \sum_{n=0}^{\infty} \sum_{m=0}^{\infty}A_{n,m}\sinh((n+1/2)\xi)P_n^{m}(\cos\eta)\cos(m\phi).
    \end{equation}
    In contrast, the solute flux vanishes at the wall for the said self-diffusiophoresis problem, leading to the condition $A_{n,m}=0.$
    Secondly, in contrast to concentration profile for the corresponding self-diffusiophoresis problem, here a complete description of the temperature profile demands a concurrent solution of the inner phase temperature distribution resulting from the redistribution of thermal energy inside the particle on account of the surface absorption of heat flux $ (q'') $. Thus, the physical properties of the particle material also contribute to the final temperature distribution.}
In view of the boundedness of the temperature everywhere inside the particle, the solution of the thermal field turns out to be  \citep{Subramanian2001}:
\begin{equation}
\label{eq:Temp_in_1}
T_{p} = \sqrt{\cosh \xi - \cos\eta} \sum_{n=0}^{\infty} \sum_{m=0}^{\infty}d_{n,m} e^{-(n+1/2)\xi}
P_n^{m}(\cos\eta)\cos(m\phi + \gamma_m).
\end{equation}
Additionally, the symmetry of the metal cap about the $ x$-$z $ plane gives $ \gamma_{n,m}=0$.

\subsection{Hydrodynamics of near-wall self-thermophoresis}
\label{ssec:hydro}
The selective absorption of heat along the swimmer surface creates an asymmetric distribution of temperature around the particle. Since, the length scale $ (\widetilde{l}) $ of interaction between the particle and suspending fluid is much smaller than the microswimmer radius $ (\widetilde{l} \ll a) $, the flow pattern can be obtained following the boundary layer theory \citep{Anderson1989,Golestanian2007,Wuerger2010}. The genesis of a tangential temperature gradient creates an osmotic pressure difference, which, in turn, drives a thermoosmotic slip flow along the particle surface, given as \citep{Anderson1989,Kroy2016}
\begin{equation}\label{eq:u_surf}
\mathbf{\widetilde{u}}^s=\mathcal{M}  (\mathbf{I-{n}{n}})\nabla T_s.
\end{equation}
Here $ \mathcal{M}  $ denotes the thermophoretic mobility characterizing the particle-fluid interaction and is defined as $ \mathcal{M} =-\dfrac{\widetilde{l}^2 \bar{\mathfrak{H}
}}{\mu T_0}, $ where $ \mu $ is the fluid viscosity, $ \bar{\mathfrak{H}
} $ is the characteristic value of the excess enthalpy, and $ T_0 $ is the ambient temperature. Since the fluid phase is  initially immobile in the present scenario, this thermoosmotic flow is tantamount to a corresponding phoretic movement of the particle. In the present demonstrations, we consider a negative value of the excess enthalpy $ (\bar{\mathfrak{H}
}<0) $, which results in a fluid flow directed opposite to the temperature gradient in the surrounding fluid medium \citep{Jiang2010,Weinert2008}.

Subsequently, we incorporate a non-dimensionalization scheme, where different quantities are normalized by the following reference values: length $\sim a $, heat flux $\sim q''$, thermal conductivity $ \sim \varkappa_f $, temperature $\sim q'' \, a/ \varkappa_f$, and velocity $ \sim - (\widetilde{l}^2 \, \bar{\mathfrak{H}
} \, q'' \, a/\mu \, T_0 \, \varkappa_f) $.  Accordingly, we use the dimensionless temperature $\mathcal{T}$, dimensionless velocity $\mathbf{u}$, and particle to fluid thermal conductivity ratio $ \mathcal{K}=\varkappa_p/\varkappa_f$ in the analysis.

In the creeping flow limit \citep{Happel1983}, the flow field obeys the incompressibility condition and the Stokes equation, as follows:
\begin{equation}\label{eq:creeping}
\nabla \cdot {\mathbf{u}}=0 \quad \text{and} \quad -\nabla {p}+ \nabla ^2 {\mathbf{u}}=0.
\end{equation}
We analyse the hydrodynamic problem by observing the particle swimming from the laboratory reference frame. Thus, the flow velocity obeys the following boundary condition at the particle surface \begin{equation}\label{eq:BC_swimmer}
\mathbf{u}_s=\mathbf{V}+\boldsymbol{\Omega} \times \mathbf{r}_{O'} + \mathbf{u}^s,
\end{equation}
where $\mathbf{V,\Omega}$ are the particle linear and rotational velocities, respectively. On the other hand, a no-slip boundary condition is realized at the plane wall. The problem gets further simplified in view of the assumed axisymmetric coverage of the metal cap. This restricts the particle trajectory in the plane containing the wall normal and the symmetry axis of the swimmer. At the same time, the particle can rotate along an axis oriented orthogonal to both the plane wall normal and microswimmer director. Thus, the swimmer kinematics can be fully captured by focusing on $V_x, V_z,$ and $ \Omega_y$. Additionally, due to its neutrally buoyant nature, the microswimmer will experience net-zero hydrodynamic force and torque conditions, i.e.
\begin{equation}\label{eq:force_free}
\mathbf{F}= \iint \limits_{S_p}^{} \boldsymbol{\sigma}\cdot \mathbf{n_p}\, dS=0 \quad \text{and} \quad \mathbf{C}= \iint \limits_{S_p}^{} \mathbf{r}_{O'} \times (\boldsymbol{\sigma}\cdot \mathbf{n_p})\, dS=0.
\end{equation}
Here $ S_p $ denotes the swimmer surface, $ \mathbf{n_p} $ is the unit normal to the swimmer surface, and $ \boldsymbol{\sigma} $ is the fluid stress tensor.

{Similar to the earlier works on force-free microswimming \citep{Lauga2009,Poddar2020}}, we decompose the hydrodynamic problem into two sub-problems:
\textit{(I)-the thrust problem} that depicts fluid flow around a fixed microswimmer experiencing only a thermophoretic slip at the surface and
\textit{(II)-the drag problem} dealing with the rigid body motion of a spherical particle where the hydrodynamic drag is only in action. Thus, the force and torque-free conditions can be written as
\begin{equation}\label{eq:force_bal_z}
\mathbf{F}^{(Drag)}+\mathbf{F}^{(Thrust)}=0 \quad \mathrm{and} \quad \mathbf{C}^{(Drag)}+\mathbf{C}^{(Thrust)}=0, \quad \text{respectively.}
\end{equation}
The force and torque on a particle that is translating and rotating near a wall can be obtained by following the earlier works on a passive particle motion \citep{ONeill1964,Dean1963,Pasol2005} and hence are not repeated here for the sake of brevity. Using the expressions for force and torque, determined for either unit velocity or rotation, as the resistance factors $(\boldsymbol{\mathcal{R}}_T,\boldsymbol{\mathcal{R}}_C, \text{ and } \boldsymbol{\mathcal{R}}_R $), we proceed to evaluate the drag force and torque for the unknown swimming velocity $(\mathbf{V})$ and rotation $(\boldsymbol{\Omega})$, as given below
\begin{subequations}
    \label{eq:force_torque_bal_zxy}
    \begin{equation}\label{eq:force_bal}
    \mathbf{F}^{(Drag)}=\boldsymbol{\mathcal{R}}_T\cdot \mathbf{U}+\boldsymbol{\mathcal{R}}_C^T\cdot \mathbf{\Omega},
    \end{equation}
    \begin{equation}
    \mathbf{C}^{(Drag)}=\boldsymbol{\mathcal{R}}_C\cdot \mathbf{U}+\boldsymbol{\mathcal{R}}_R\cdot \mathbf{\Omega}.
    \end{equation}
\end{subequations}
The task remains is to determine the thrust force and torque on the microswimmer due to the self-thermophoretic actuation.

    \subsection{Employing the Reciprocal theorem}
\label{ssec:reciprocal}
To obtain the propulsive thrust on the microswimmer, we use the Reynolds reciprocal theorem (RRT) applicable between two Stokes flows with similar geometry. It has the following general form \citep{Happel1983}:
\begin{equation}\label{eq:reciprocal_gen}
\iint_{\partial S} \mathbf{{n}} \cdot \boldmath{\sigma '} \cdot \mathbf{u ''} = \iint_{\partial S} \mathbf{{n}} \cdot \boldmath{\sigma ''} \cdot \mathbf{u '}.
\end{equation} 
Here ` $ ' $ ' and ` $ '' $ ' - superscripted variables correspond to those associated with the thrust problem and a complementary Stokes problem, respectively;  $ \partial S $ denotes the fluid confining boundary. Keeping in view of  the no-slip condition for fluid velocity at the plane wall and decaying flow field at large distances from the microswimmer, we can simply use the swimmer surface $ (S_p) $ in place of  $ \partial S $ in \eqref{eq:reciprocal_gen}.
Choosing the complementary Stokes problem as the motion of a spherical particle near a plane wall with the magnitude of velocity to be unity ($ \{\mathbf{U''},\boldsymbol{\Omega''}\}=\{\mathbf{e}_x+\mathbf{e}_z, \mathbf{e}_y\})$ and utilizing the boundary condition on the  surface of a fixed swimmer (\eqref{eq:BC_swimmer}), the reciprocal relation (\eqref{eq:reciprocal_gen}) gives the thrust components on the swimmer as     
\begin{equation}\label{eq:reciprocal_swimmer}
{F^{(Thrust)}_x}=\iint_{S_p} \mathbf{{n}} \cdot \boldsymbol{ \sigma_{A}''} \cdot \mathbf{u^s}\,dS,  \quad {F^{(Thrust)}_z}=\iint_{S_p} \mathbf{{n}} \cdot \boldsymbol{ \sigma_{B}''} \cdot \mathbf{u^s}\,dS ,
\quad \text {and} \;\;
{C^{(Thrust)}_y}=\iint_{S_p} \mathbf{{n}} \cdot \boldsymbol{ \sigma_{C}''} \cdot \mathbf{u^s}\,dS.
\end{equation}
 Here suffixes A, B and C on the stress tensors denote the corresponding fundamental problems of a translating and rotating particle in different directions, reported previously by \citet{ONeill1964,Pasol2005} and \citet{Dean1963}, respectively.

Although evaluated at the particle surface, the stress tensors in the force and torque expressions of \eqref{eq:reciprocal_swimmer} implicitly contain information about the flow boundary condition at the wall, i.e. the no-slip and no-penetration flow  conditions. On the other hand, the surface flow velocity embodies the effect of the specific thermal boundary condition at the wall due to the relation in \eqref{eq:u_surf}.  The thermal and the hydrodynamic field distributions are further interconnected through the kinematic constraints imposed by the wall and the constraints on force and torque (\eqref{eq:force_free}).

\section{Results and discussion}

In the experimental realizations of the auto-thermophoresis \citep{Jiang2010,Qian2013,Chen2018}, materials that have been used so far to fabricate a thermally asymmetric particle include microspheres made of ceramics, such as fused silica; and polymers, such as polystyrene, while the continuous fluid medium has been commonly taken as a mixture of water and glycerol. The typical thermal conductivities of these materials are given by: 1.3 W/mK - fused silica \citep{Touloukian1970}, 0.13 W/mK - polystyrene \citep{Sombatsompop1997}, and 0.54 W/mK - water-glycerol mixture \citep{Association1963}. Considering these materials, the thermal conductivity ratio becomes $ \mathcal{K}=0.24 $ (polystyrene-water) or $ \mathcal{K}=2.407 $ (silica-water). However, a plenty of other materials have also been used to produce Janus particle with different functionalities \citep{Hu2012} and are yet to be tested for their performance in auto-thermophoresis.
It should be noted that, in the previous theoretical treatments of an unconfined self-thermophoresis \citep{Jiang2010,Bickel2013}, the parameter $ \mathcal{K} $ has simply been chosen as 1  considering the  same order of magnitudes of the particle and fluid thermal conductivities. Thus, remaining consistent with the practical values and from the theoretical interest of capturing the key physical aspects of the thermal conductivity variation of the particle or fluid, we  vary the thermal conductivity ratio $ (\mathcal{K}) $ from 0.1 to 10 during the illustration of results.
\subsection{Temperature profile}
\label{ssec:temp_result}
\begin{figure}[!htb]
    \centering
    \includegraphics[width=0.85\textwidth]{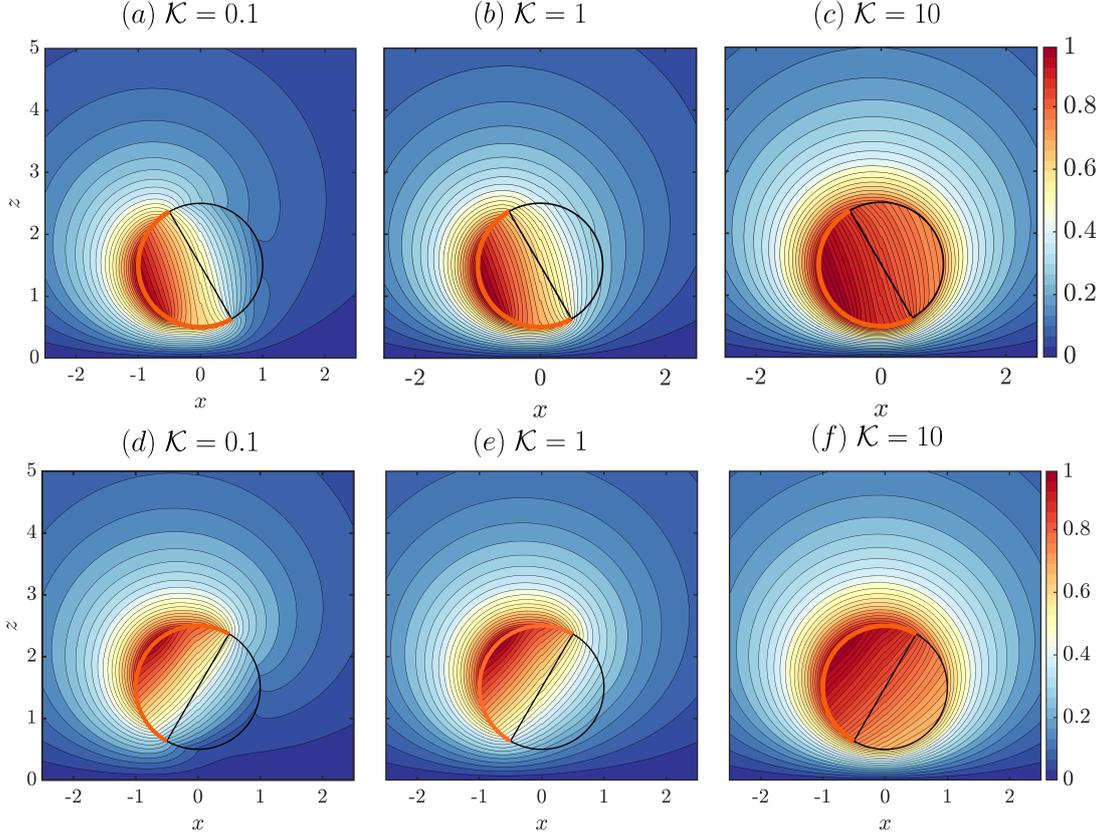}
    \caption[Spatial variation of the scaled temperature $ (\bar{\mathcal{T}}=T/(T_{max}-T_{min})) $ in the $x$-$z$ plane, both inside the microswimmer and in the surrounding fluid medium, for a fixed wall-swimmer distance of $ \delta=0.5$ and a coverage angle of the metal coating $ \varphi_\text{cap}=90^{\circ} $. ]
    {Spatial variation of the scaled temperature $ (\bar{\mathcal{T}}=T/(T_{max}-T_{min})) $ in the $x$-$z$ plane, both inside the microswimmer and in the surrounding fluid medium, for a fixed wall-swimmer distance of $ \delta=0.5$ and a coverage angle of the metal coating, $ \varphi_\text{cap}=90^{\circ} $. The orientation angle ($ \theta_{p} $) is $ 60^{\circ}$ in (a)-(c), while it is $ 120^{\circ}$ in (d)-(f). The thermal conductivity contrast has been varied as 0.1, 1 and 10 from left to right panels. The thick orange arc in each figure represents the metal-coated area of the surface.}
    \label{fig:temp_compact_60}
\end{figure}

\begin{figure}[!htb]
    \centering
    \includegraphics[width=1\textwidth]{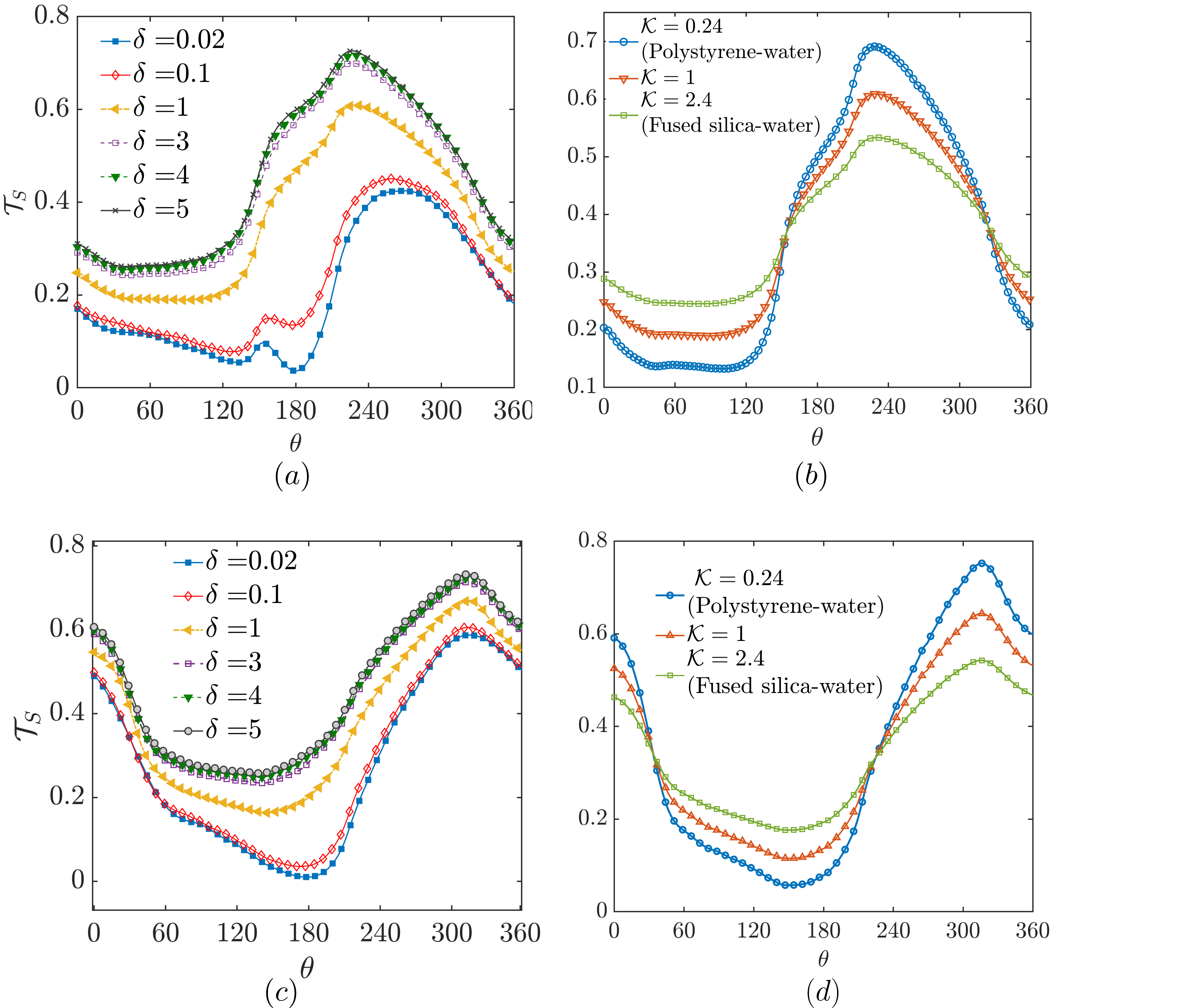}
    \vspace{-3ex}
    \caption[(a),(b): Surface temperature $ (\mathcal{T}_S) $ along the surface for different wall distances and thermal conductivity ratios, respectively. (c),(d): Similar variations for a different inclination, $ \theta_{p} = 120 ^{\circ} $.]
    { (a),(b): Surface temperature $ (\mathcal{T}_S) $ variation along the surface for different wall distances and thermal conductivity ratios, respectively. Here, the angle $ \theta $ is at the surface of the microswimmer, measured clockwise from positive $ z $ direction. In (a), we have taken $ \mathcal{K}=1 $, while in (b) $ \delta=1$. Other parameters are $ \varphi_\text{cap} =90 ^{\circ}$ and $ \theta_{p} =60 ^{\circ}$. In (c),(d), similar variations are shown for a different inclination of $ \theta_{p} = 120 ^{\circ} $. } \label{fig:temp_compact_line}
\end{figure}

In figure~\ref{fig:temp_compact_60}, we describe the effects of certain parameters on the resulting temperature profile due to self-thermophoresis. In the first case (figures~\ref{fig:temp_compact_60} (a)-(c)), the microswimmer director is leaning towards the wall, i.e. the cold portion is facing away from the isothermal wall $ (\theta_p=60^{\circ})$. Similarly, in figures~\ref{fig:temp_compact_60} (d)-(f), the inclination angle is $ 120^{\circ}$. The presence of the wall breaks the symmetry of the temperature distribution along the director axis.
$ \mathcal{T}_S $ is highly dependent on its distance from the wall, as shown in figures~\ref{fig:temp_compact_line}(a) and (c). In these figures, the variation of the angle $ \theta $ denotes the distance along the surface starting from the $ + \text{ve} \; z$ axis. {Sensing a nearby low-temperature thermal obstacle, the micromotor surface temperature drops when the swimmer approaches more towards the wall, irrespective of its inclination.}
Besides, the locations of the maximum and minimum temperatures on the particle surface vary with the distance. For both of the inclinations considered, the minimum temperature shifts more towards the wall as the swimmer enters a wall-adjacent zone. In the former configuration, where the heated portion is more close to the wall, the maximum temperature gets a shift away from the wall as $ \delta $ decreases. However, in the latter configuration, the maximum temperature location has only a negligible shift in the same direction.

The above-mentioned maxima and minima locations of the surface temperature control the direction of surface slip flow, which in the present case, is directed from a colder point to a hotter one on the surface. 

Comparing the temperature distributions in figures~\ref{fig:temp_compact_60} (a) to (c) or (d) to (f) we find that, as the particle material becomes less and less thermally conductive than that of the surrounding fluid $ (\mathcal{K}<1) $, the hot-spot on the particle surface becomes progressively more localised. In the limit of $ \varkappa_p \ll \varkappa_f $ or $ \mathcal{K} \ll 1 $, the surface of the metal-coated portion shows a conducting nature, while the bare polymeric surface behaves as a perfect insulating body. This is the case when the extent of temperature asymmetry around the micromotor reaches its maximum. Similarly, in the above figures, for $ \mathcal{K}=0.1 $ and 1, the metal-coated and uncoated halves show a vivid contrast in the isotherm contours coming out from the surface of the particle.
On the other hand, with $ \varkappa_{p} $ much higher than $\varkappa_{f} $, the hot area occupies a greater extent of the particle.
However, in the limit of $ \mathcal{K} \gg 1 $, the particle becomes highly conducting and behaves almost as an isothermal body. This trend is reflected in figures~\ref{fig:temp_compact_60} (c) and (f), where the fluid medium isotherms cut the surface almost parallel to it, and the insulating nature of the uncoated half is merely observed. Hence, even with a low cap angle, the particle can behave similarly to the case of high cap coverage if the particle thermal conductivity is high.
The effect of thermal conductivity contrast between the particle and fluid is also prominently extended in the fluid domain. For example, in the case of a highly conducting particle, the heated zone in the fluid almost surrounds the full circumference of the particle, as portrayed in figures~\ref{fig:temp_compact_60}(c) and (f).

The effects of contrasting thermal conductivities on the surface temperature profile have been demonstrated in figures~\ref{fig:temp_compact_line} (b) and (d) for inclinations $ 60^{\circ} $ and $ 120^{\circ} $, respectively. We have chosen typical values of $ \mathcal{K} $ for which the particle is more conductive than the fluid ($\mathcal{K}>1$) and vice versa ($ \mathcal{K}<1 $). With the hot surface facing the wall, the $ \mathcal{K}>1 $ condition causes a raise in the surface temperature before reaching $ \theta \approx 155^ \circ $; thereafter the temperature decays until $ \theta \approx 325^ \circ $ and again raises beyond this location. Locations of these turnover points on the surface as well as the behaviour of temperature with changing $ \mathcal{K} $ between these points are modified when the cold portion faces the wall.

\subsection{Effect of thermal conductivity contrast in an unbounded flow}
\label{ssec:vel_ub}
\begin{figure}[!htb]
	\centering
	\begin{subfigure}{0.65\textwidth}
		\centering
		\includegraphics[width=1\textwidth]{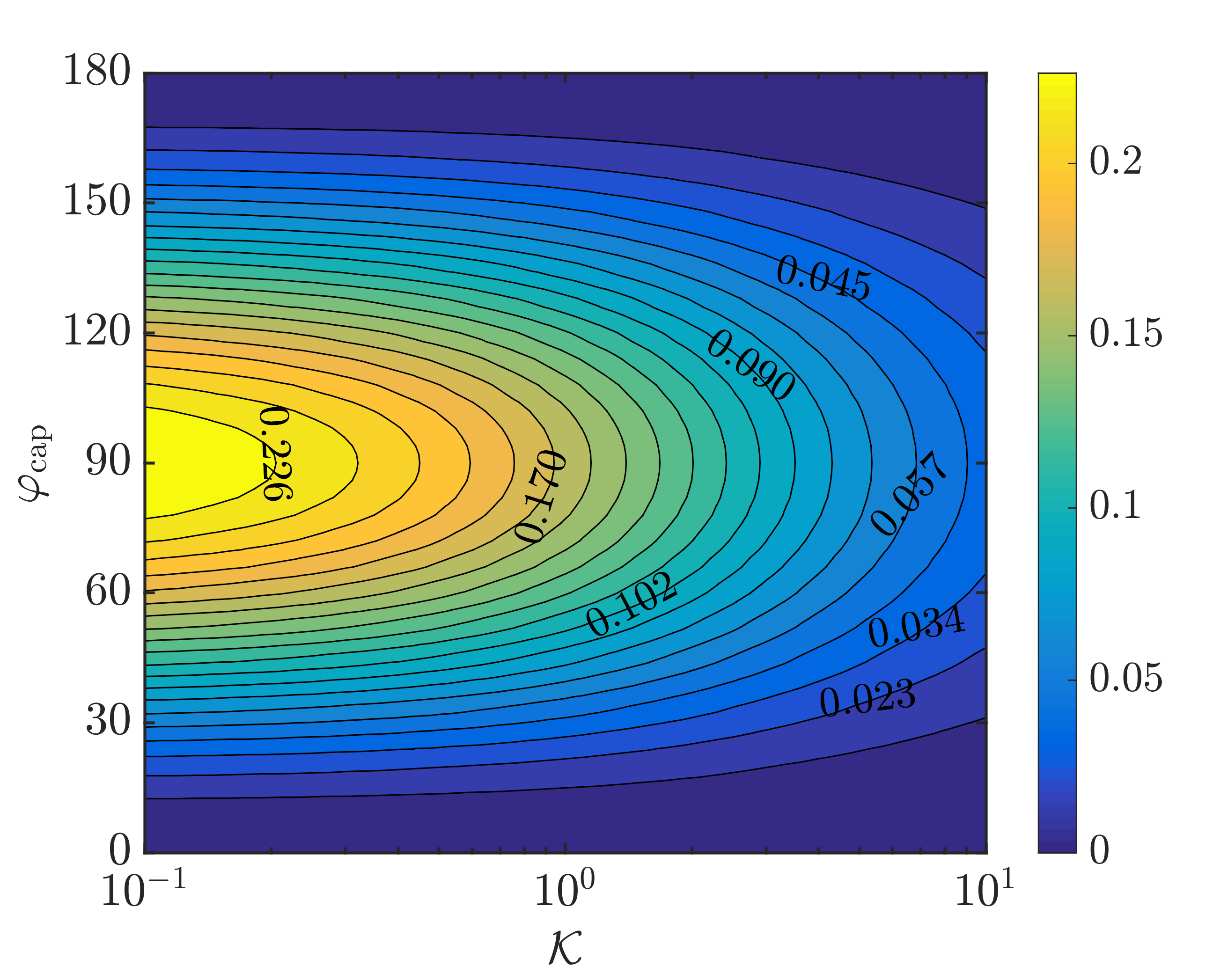}
		\vspace{15ex}
		\caption{}
		\label{fig:contour_Vz_tcap_K}
	\end{subfigure}
\\
	\begin{subfigure}{0.47\textwidth}
		\centering
		\vspace{3.5ex}
		\includegraphics[width=1.1\textwidth]{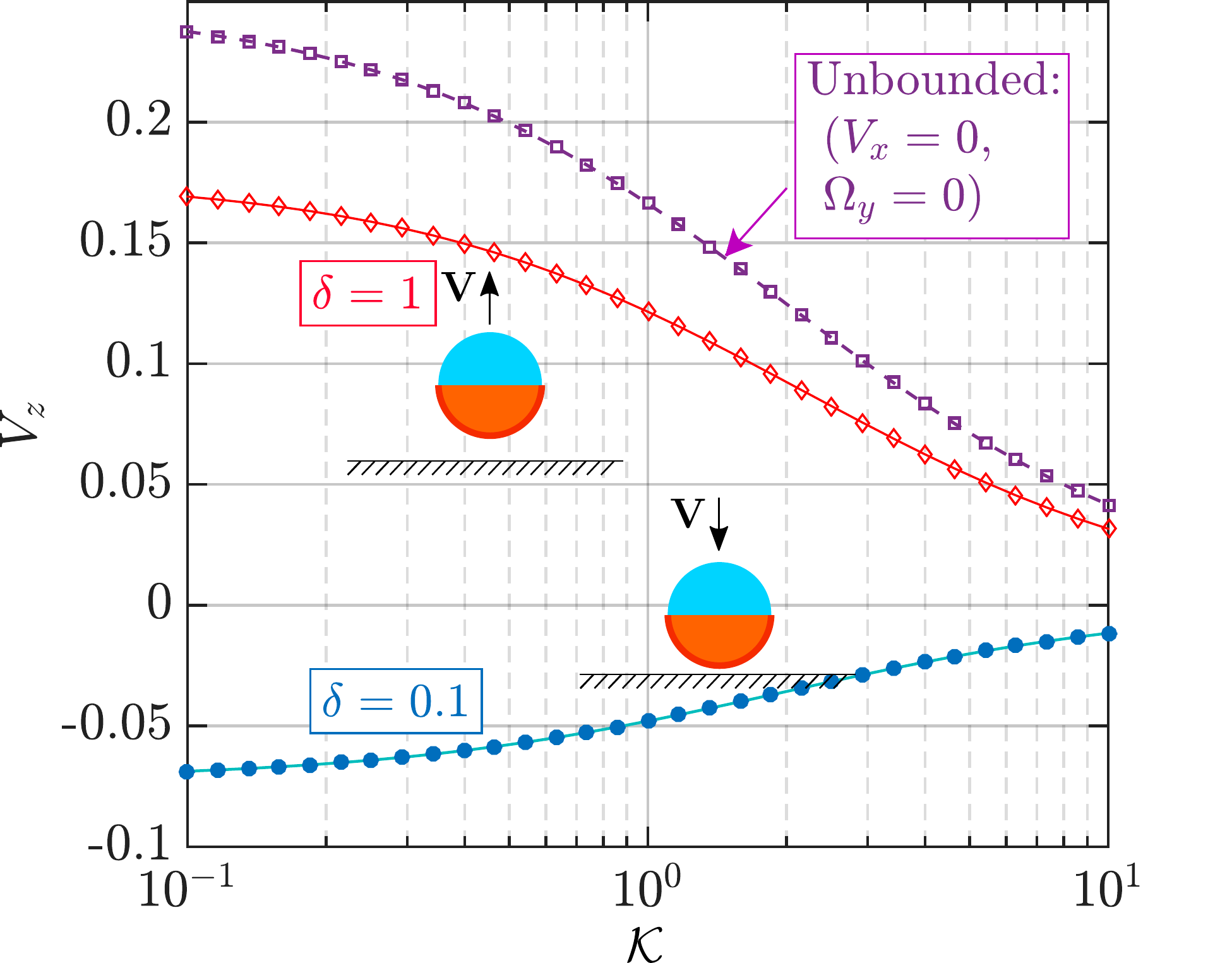}
		\vspace{9.5ex}
		\caption{}
		\label{fig:vz_vs_K_vary_delta_tp_0_tcap_90}
	\end{subfigure}
\qquad \quad
	\begin{subfigure}{0.44\textwidth}
		\centering
		\includegraphics[width=1.12\textwidth]{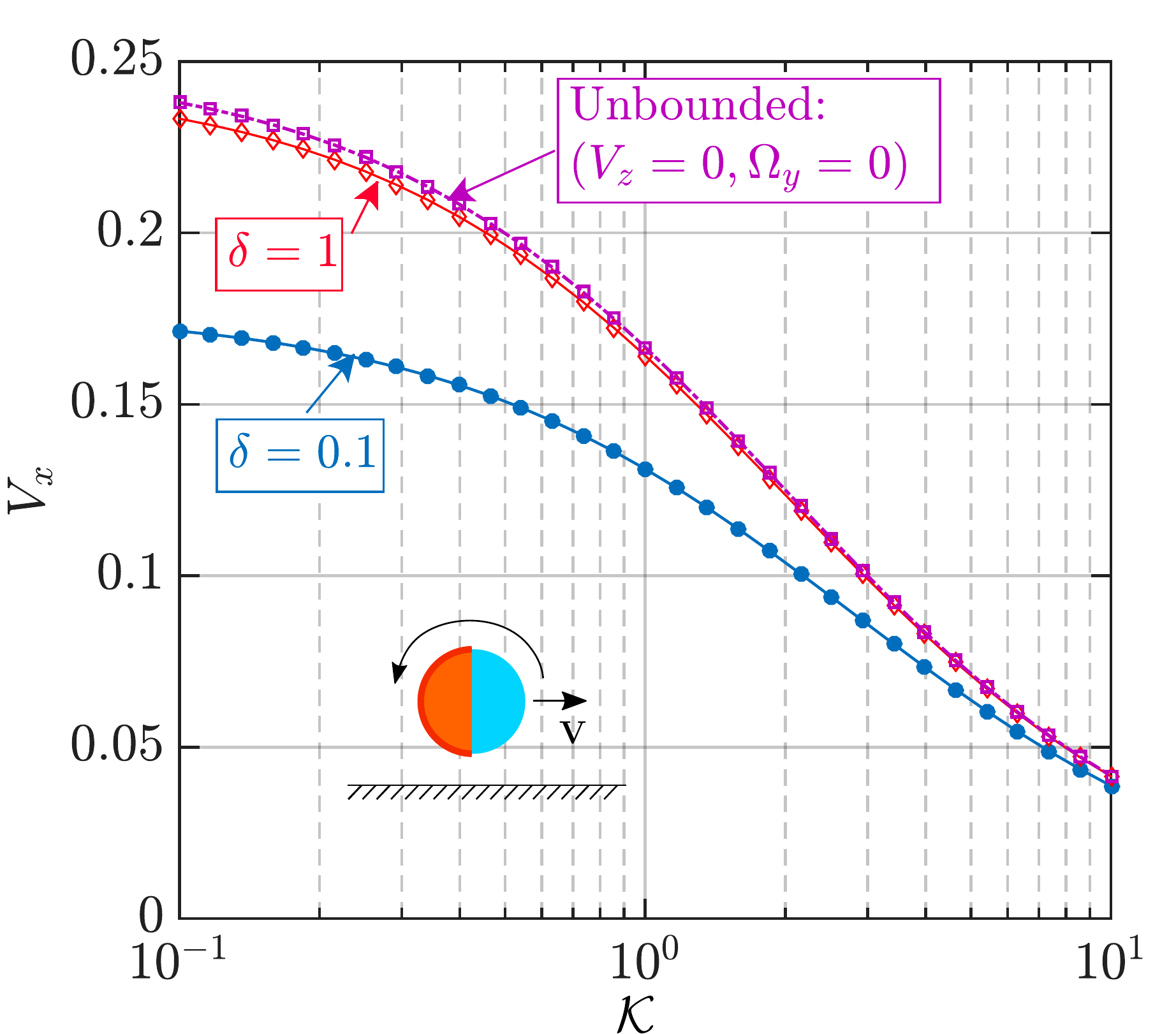}
		\vspace{7.5ex}
		\caption{}
		\label{fig:vx_vs_K_vary_delta_tp_90_tcap_90}
	\end{subfigure}
	   \vspace{1ex}
	\caption[]
	{In (a), the unbounded swimming velocity magnitude $ |\mathbf{V}_\text{ub}| $  is shown in the $ \mathcal{K}-\varphi_\text{cap}$ plane.   In (b) and (c), the variations in velocity components ($V_z$ and $V_x$, respectively) with the thermal conductivity contrast ($\mathcal{K}$) are shown for both unbounded and near-wall scenarios. The orientation and velocity direction of the particle are shown schematically in each figure.}
	\label{fig:vel_ub_compare}
\end{figure}
\textcolor{black}{Before investigating the coupled interplay between the thermal conductivity contrast and the bounding wall  on the microswimmer velocity, we delineate its behaviour in an unbounded domain.
In an unbounded flow, the rotational component of the swimmer velocity does not exist since the wall-induced asymmetry in the temperature and flow field are absent in this case. The linear velocity magnitude $|\mathbf{V}_\text{ub}|=\sqrt{V_x^2+V_z^2}$ depends only on the  metallic cap coverage, $\varphi_\text{cap}$ and the thermal conductivity ratio, $\mathcal{K}$, as portrayed in 
figure~\ref{fig:contour_Vz_tcap_K}. It is observed that the effect of $\varphi_\text{cap}$ on $|\mathbf{V}_\text{ub}|$ is symmetric about the $50 \% $ coverage $(\varphi_\text{cap}=90^{0})$. On the other hand, the effect of increasing $\mathcal{K}$ is to decrease the velocity magnitude due to reduced temperature gradients.}

\textcolor{black}{The wall-induced distortion of the temperature field weakens asymptotically and beyond a certain height, the thermal boundary condition at the wall is likely to become inconsequential.  A simultaneous decay in the hydrodynamic disturbance also 
takes place, tending towards a situation equivalent to that of an unbounded scenario. We quantify this distance with an effective separation distance, $\delta_\text{ub}$, at which a velocity component of the microswimmer becomes $99 \% $ of the unbounded velocity magnitude $|\mathbf{V}_\text{ub}|$.
This effective distance is a function of the thermal and configurational parameters, i.e. $\delta_\mathrm{ub}=\delta_\text{ub}(\mathcal{K},\theta_p,\varphi_\text{cap})$.
As an example of this functional dependence, we show two reference configurations of the particle motion in 	\ref{fig:vz_vs_K_vary_delta_tp_0_tcap_90} and \ref{fig:vx_vs_K_vary_delta_tp_90_tcap_90}, where either of the linear velocity component of the microswimmer exists in an unbounded domain. While in the first instance (figure~\ref{fig:vz_vs_K_vary_delta_tp_0_tcap_90}), the $\delta_\text{ub}$ for vertical velocity component reduces from $\delta_\text{ub}=7.58$ for $\mathcal{K}=0.1$ to $\delta_\text{ub}=6.89$ for $\mathcal{K}=10$, 
in the second instance  (figure~\ref{fig:vx_vs_K_vary_delta_tp_90_tcap_90}) the $\delta_\text{ub}$ for horizontal velocity component reduces from $\delta_\text{ub}=1.61$ for $\mathcal{K}=0.1$ to $\delta_\text{ub}=0.63$ for $\mathcal{K}=10$. The difference in the $\delta_\text{ub}$ for different velocity components indicates a contrasting nature of the wall effects in different flow directions.}
\subsection{Combined interplay of bounding wall and thermal conductivity contrast}
\label{ssec:vel_result}
\begin{figure}[!htb]
    \centering
    \begin{subfigure}{0.44\textwidth}
        \centering
        \includegraphics[width=1\textwidth]{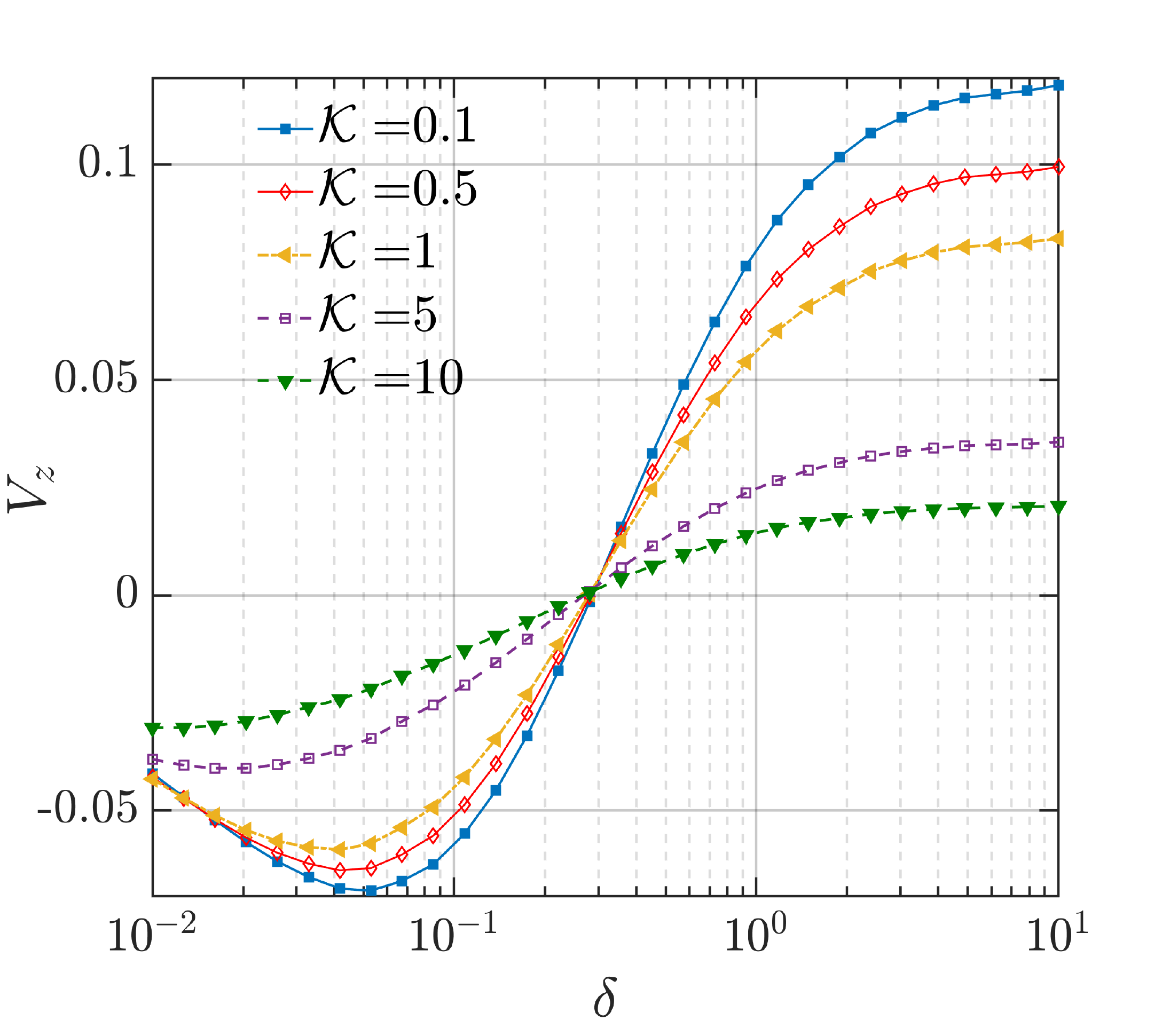}
         \vspace{7.5ex}
        \caption{}
        \label{fig:Vz_vary_delta_vary_K_tc_90_tp_60}
    \end{subfigure}
    \begin{subfigure}{0.45\textwidth}
        \centering
        \includegraphics[width=1\textwidth]{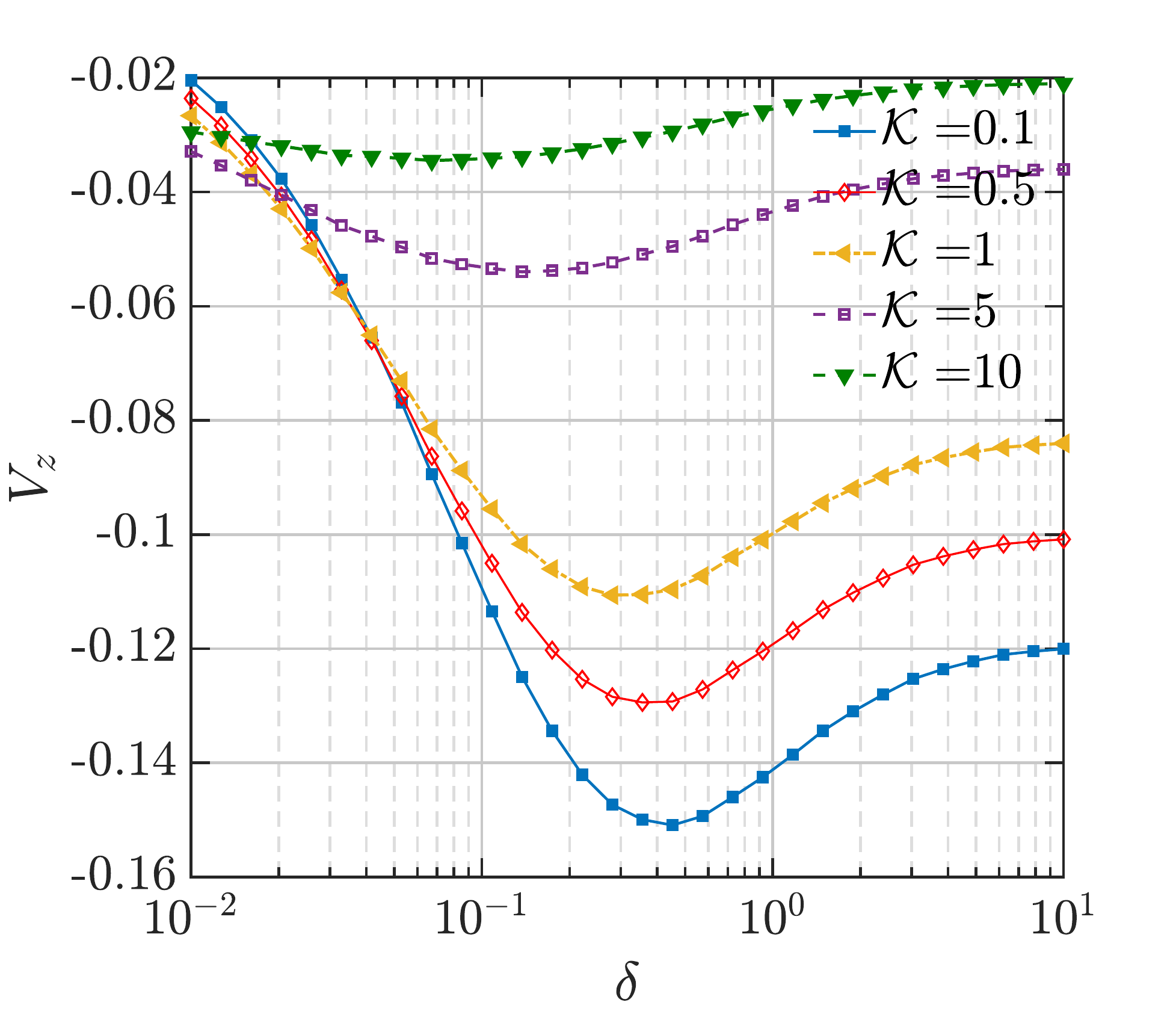}
         \vspace{7.5ex}
        \caption{}
        \label{fig:Vz_vary_delta_vary_K_tc_90_tp_120}
    \end{subfigure}
    \qquad
    \begin{subfigure}{0.43\textwidth}
        \centering
        \includegraphics[width=1.1\textwidth]{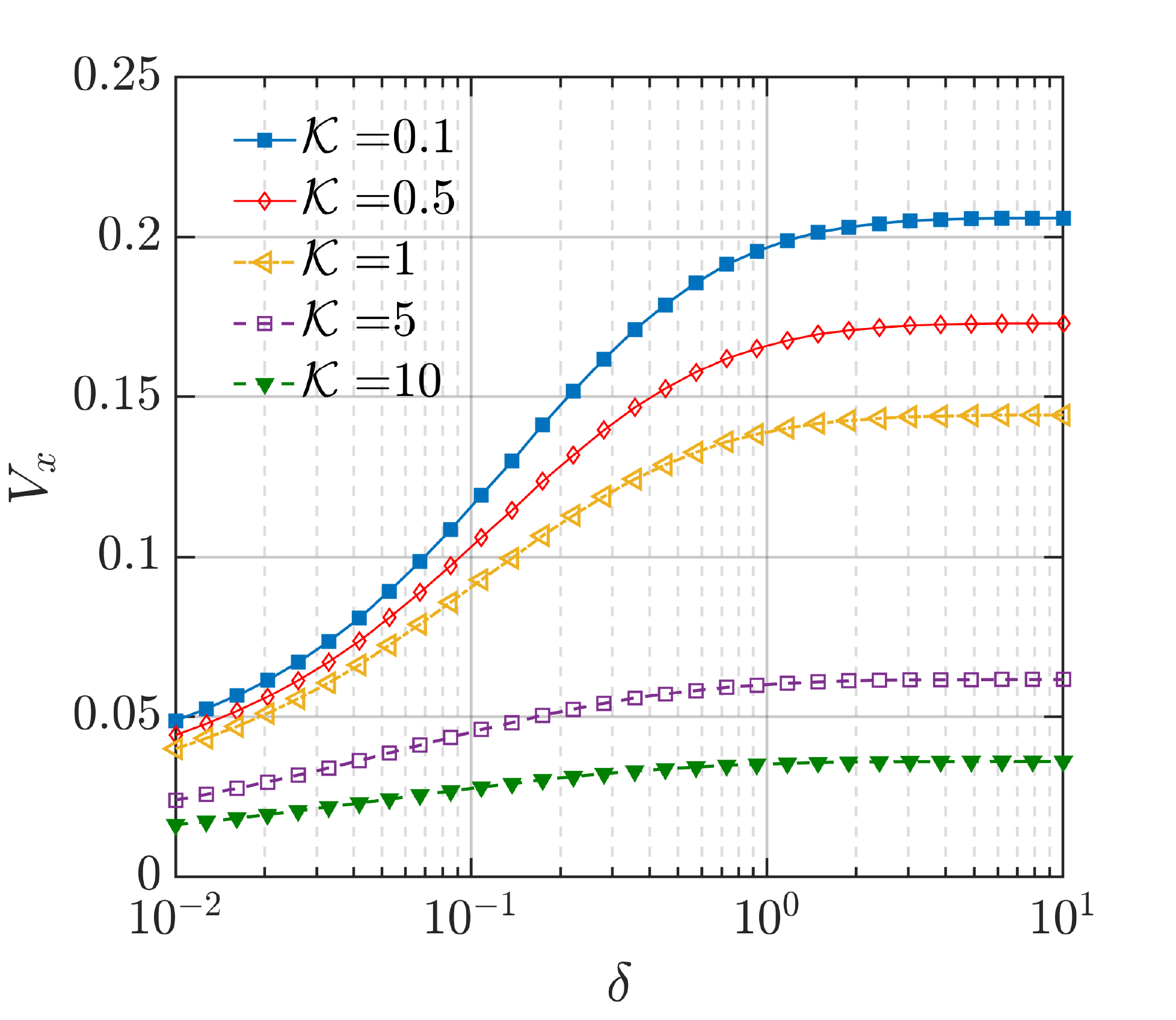}
         \vspace{7.5ex}
        \caption{}
        \label{fig:Vx_vary_delta_vary_K_tc_90_tp_120}
    \end{subfigure}
    \begin{subfigure}{0.43\textwidth}
        \centering
        \includegraphics[width=1.1\textwidth]{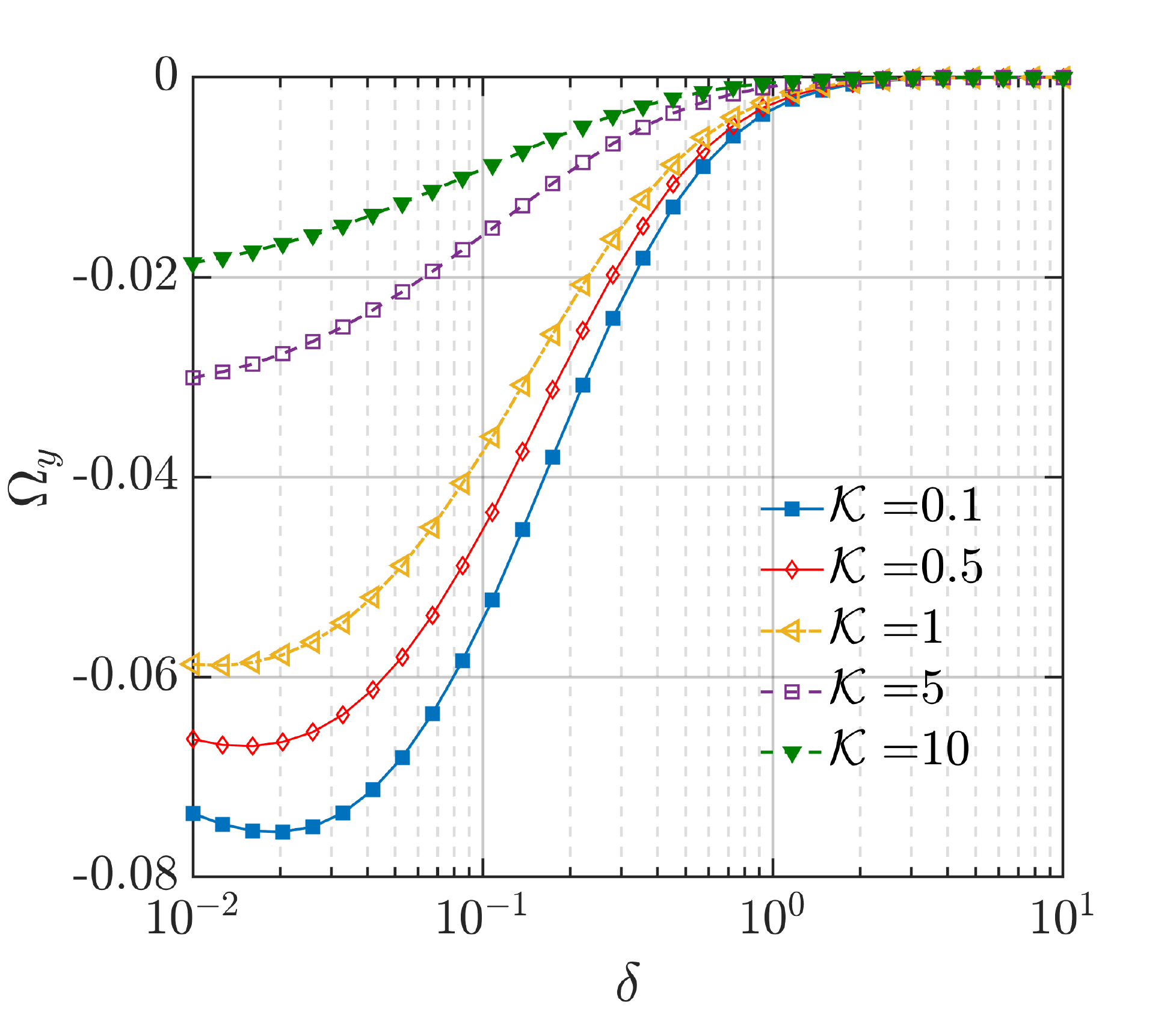}
         \vspace{7.5ex}
        \caption{}
        \label{fig:Wy_vary_delta_vary_K_tc_90_tp_120}
    \end{subfigure}
    \caption[Variation of translational and rotational velocity components of the microswimmer with distance from the wall $ (\delta) $ and various thermal conductivity ratios $ (\mathcal{K}) $.]
    {Variation of translational and rotational velocity components of the microswimmer with distance from the wall $ (\delta) $ and various thermal conductivity ratios $ (\mathcal{K}) $. The heated cap coverage is identically chosen as $ \varphi_\text{cap} =90 ^{\circ} $ for all the cases. In subfigures (a) and (b), $ \theta_{p} =60 ^{\circ}$ and $120 ^{\circ}$, respectively. Due to symmetry reasons, the subfigures (c) and (d) are applicable for both $ \theta_{p}= 60 ^{\circ}$ or $120 ^{\circ}$.}
    \label{fig:vel_comps}
\end{figure}
\begin{figure}[!htb]
    \centering
    \begin{subfigure}{0.43\textwidth}
        \centering
        \includegraphics[width=1.1\textwidth]{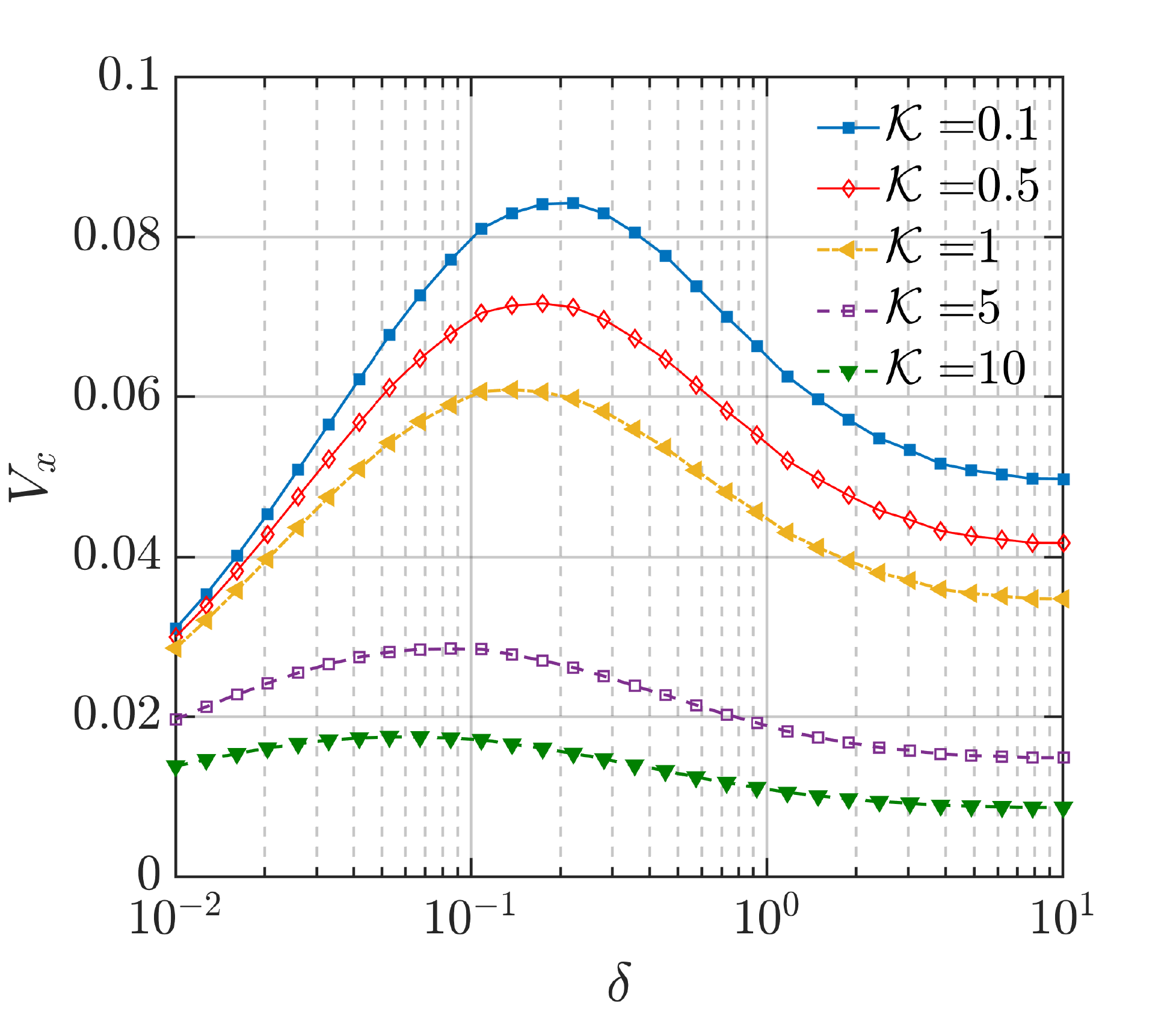}
         \vspace{5ex}
        \caption{}
        \label{fig:Vx_vary_delta_vary_K_tc_140_tp_150}
    \end{subfigure}
    \quad
    \begin{subfigure}{0.43\textwidth}
        \centering
        \includegraphics[width=1.1\textwidth]{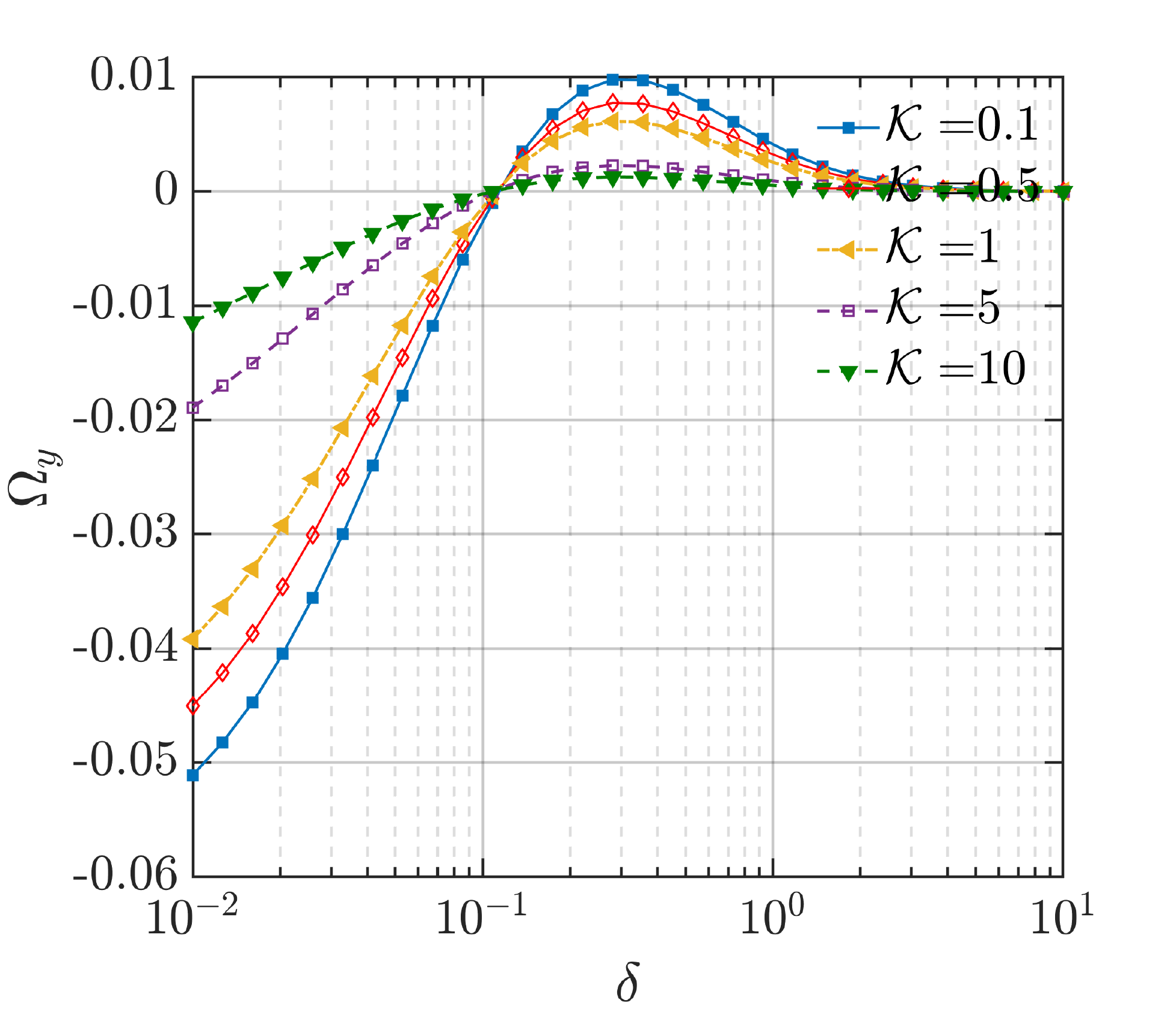}
         \vspace{5ex}
        \caption{}
        \label{fig:Wy_vary_delta_vary_K_tc_140_tp_150}
    \end{subfigure}
    \caption[Variation of wall-parallel translational and rotational velocities of the microswimmer with distance from the wall $ (\delta) $ and different thermal conductivity ratios $ (\mathcal{K}) $. ]
    {Variation of wall parallel translational and rotational velocities of the microswimmer with distance from the wall $ (\delta) $ and different thermal conductivity ratios $ (\mathcal{K}) $. Other parameters are chosen as $ \varphi_\text{cap} =140 ^{\circ}$ and $ \theta_{p} =150 ^{\circ}$.}
    \label{fig:vel_comps_2}
\end{figure}

In both the figures~\ref{fig:Vz_vary_delta_vary_K_tc_90_tp_60} and \ref{fig:Vz_vary_delta_vary_K_tc_90_tp_120}, the downward translation of the swimmer is intensified due to the presence of the wall, and subsequently it reaches an optimum at a certain vertical distance $ (\delta_{opt})$. With further separation from the wall, this motion is retarded. When the coated surface faces the wall $ (\theta_p=60^\circ) $, $V_z$ remains negative till a certain distance of $\delta_{cr} \approx 0.3$ from the wall; then it shows a trend of positive $V_z$. Finally, the swimmer gradually reaches a velocity that it would have attained if isolated.
In the second configuration $(\theta_p=120^\circ)$, where the uncoated surface is nearer to the wall, the swimmer continues to translate downwards irrespective of the wall distance.
For both the configurations, the location of $\delta_{opt}$ is shifted far from the wall, as the particle becomes more and more conductive (increasing $\mathcal{K}$). Such a consequence implies the coupled interplay between the particle-to-fluid thermal conductivity ratio and the wall-distance in influencing the particle velocity.

The vertical translation of the particle is  decoupled from the other translational or rotational components. Hence, the velocity reversal in this direction exists solely due to the vertical phoretic thrust exerted on the particle. As previously discussed, a variation in the swimmer-wall distance intervenes with the swimmer surface temperature distribution, and thereby its gradient along the surface is also affected. Because of this, the slip flow at the particle surface (\eqref{eq:u_surf}) is modified. Simultaneous to the distortions in the temperature field, the hydrodynamic stress distribution around the particle is also disturbed by the confining boundary. Both of these mechanisms work behind an altered thrust force experienced by the microswimmer (\eqref{eq:reciprocal_swimmer}).

Following the same figures, the thermal conductivity contrast $ (\mathcal{K})$ can neither shift the critical distance for the velocity reversal nor it can alter the sign of velocity in both the swimmer inclinations considered.
However, for a specific wall to swimmer distance $ (\delta) $, the swimmer translates slower with the escalating values of $ \mathcal{K}$. While uncovering the associated physical mechanism, we refer to the attenuation of the asymmetry in temperature around the coated and uncoated faces of the microswimmer with high particle conductivity (see figures~\ref{fig:temp_compact_60}(c) and (f)), causing attenuation in the surface temperature gradient. This occurrence can be quantitatively visualised from the fact that, in figures~\ref{fig:temp_compact_line}(b) and (d), with increasing $ \mathcal{K}$, the maximum temperature falls, but the minimum temperature hikes, despite their corresponding locations on the surface being hardly affected.
Hence, a weaker surface temperature gradient (i. e. $\mathcal{T}_{max}-\mathcal{T}_{min}$) is generated, and eventually, the surface flow is weakened. Accordingly, a diminishing surface traction strength on the swimmer surface results.
It is to be noted that, in figure~\ref{fig:Wy_vary_delta_vary_K_tc_90_tp_120}, very close to the wall $( \delta \lesssim 0.05)$, the variation in $ V_z $ shows a non-monotonic dependence on the thermal conductivity contrast, $ \mathcal{K}$. In the narrow gap region, strong hydrodynamic stresses build up, which are critically coupled with the flow modulations due to variations in the thermal conductivity. Accordingly, the thrust force varies non-monotonically with $ \mathcal{K} $, as portrayed in figure~\ref{fig:Fz_thrust_vary_K_tp_120_tc_90}.

In figures~\ref{fig:Vx_vary_delta_vary_K_tc_90_tp_120} and \ref{fig:Wy_vary_delta_vary_K_tc_90_tp_120}, we portray the variations in wall-parallel translational and rotational velocities with the parameters $ \delta $ and $ \mathcal{K}.$ The particle is slowed down as it approaches the wall, and it moves faster with lessened particle thermal conductivities. The near-wall rotation remains in the clock-wise direction throughout, although the magnitude decays with strengthened particle thermal conductivity.
While exploring other interesting facets of the velocity components, we have shown the variations in $ V_x $ and $ \Omega_y $ in figure~\ref{fig:Vx_vary_delta_vary_K_tc_140_tp_150} and \ref{fig:Wy_vary_delta_vary_K_tc_140_tp_150}, respectively for a different coverage angle of $\varphi_\text{cap} =140^{\circ}$ and inclination, $ \theta_{p} =150 ^{\circ} $. In this case, upon increasing its distance from the wall, the particle initially migrates horizontally at a faster pace, but is subsequently retarded beyond a specific separation height. Further, as the particle thermal conductivity reduces, the location of this maximum point shifts far from the wall. We also observe a change in the direction of the particle rotation from counter-clockwise to clock-wise at a fixed distance from the wall $ (\delta \approx 0.11) $.

{For gaining an insight into the physical origin of such behaviours,} we first look into the expressions of $ V_x $ and $ \Omega_y$.
{Unlike the vertical motion, the parallel rotation and translational are coupled to each other. The final translational and rotational velocities take the following forms:
    \begin{subequations}\label{eq:Vx_Wy_coupled_temp}
        \begin{gather}
        V_x=\dfrac{{F_x^\text{(Thrust)} f_{y}^R}- C_y^\text{(Thrust)} f_{x}^R }{f_{x}^Rf_{y}^T-f_{x}^T f_{y}^R} \quad \text{and} \quad \Omega_y=\dfrac{C_y^\text{(Thrust)} f_{x}^T- F_x^\text{(Thrust)} f_{y}^T}{f_{x}^Rf_{y}^T-f_{x}^Tf_{y}^R}, \tag{\theequation a-b}
        \end{gather}
\end{subequations}}
where different hydrodynamic resistance factors are defined as $F_{x}^{(Drag,T)}=f_{x}^T\,V_x,$ $F_{x}^{(Drag,R)}=f_{x}^R\,\Omega_y, \; C_{y}^{(Drag,T)}=f_{y}^T\;V_x, \;$ and $\;C_{y}^{(Drag,R)}=f_{y}^R\;\Omega_y$.
It can be inferred from equation~\ref{eq:Vx_Wy_coupled_temp}(b) that the actual sign of the rotational velocity is a complex function of the hydrodynamic resistance factors as well as the thrust force and torque.
From the previous investigations, it is known that these resistance factors are enhanced with a gradual descent of the particle towards the wall
\citep{Goldman1967,Chaoui2003}. Since the hydrodynamic resistance term in the denominator always remains negative (see figure~\ref{fig:terms_torque_tp_150_tc_140_K_0p1}), it is evident that the sign of $ \Omega_y $
is set by the relative importance of the phoretic torque  ($ C_y^\text{(Thrust)} f_{x}^T $) and the phoretic force $ (F_x^\text{(Thrust)} f_{y}^T) $, adjusted by appropriate units. From a physical perspective, these two effects correspond to the counter-clockwise torque due to the propulsive effects and the clockwise torque due to its forward movement $ (+V_x)$, respectively. Again, the counter-clockwise torque is a net effect of the differently directed surface flows based on the locations of peak points in the surface temperature (a diagrammatic representation of a similar scenario has been provided later in figure~\ref{fig:tcap_var_tmax_loc}(d)).

Inspection of the temperature profiles at different distances from the wall reveals that the minimum temperature shifts more towards the wall with a decreasing gap, while the maximum point remains almost unaffected. Subsequently, a greater strength of the clockwise surface flow and an ensuing counter-clockwise torque result.
Also, as the micromotor approaches closer to the wall, it is subjected to the elevated hydrodynamic stresses because of near-wall velocity gradients. Such effects result in the relative dominance of the counter-clockwise torque surpassing the opposing clockwise torque. Thus, at the reduced distances from the wall, we have a negative sign of $ \Omega_y$. However, beyond a critical height, the effect of increasing $V_x$ (see figure~\ref{fig:Vx_vary_delta_vary_K_tc_140_tp_150}) becomes dominant and so does the corresponding clockwise torque.

\subsection{Swimming trajectories}
\label{ssec:trajectory}
The near-wall movement of the microswimmer has been captured in the present study by considering only the deterministic forces acting on the swimmer and neglecting any stochastic contribution due to thermal fluctuation to the translational and rotational diffusion. 
\textcolor{black}{While deriving the governing differential equations, we have assumed a quasi-steady-state condition and neglected any rotational diffusion associated due to thermal fluctuations. The validity of these assumptions can be justified by considering the fact that the  ensuing diffusive transport behaviour due to thermal fluctuations becomes  important only when \citep{Golestanian2007} $a > \left( \dfrac{k_B T_\text{ref}}{\mu \, \widetilde{u}_\text{drift}} \right)^{1/2}.$
 With a practical value of the drift velocity as $\widetilde{u}_\text{drift} \sim 10 \; \mathrm{\mu m/s}$ in water, the above conditions suggests that the  particle radius $ (a) $ has to be greater than a few $100 \; \text{nm}$ for negligible enhancement in diffusive transport, consistent with the present analysis with the assumption of $a \geq 1 \mu m$.}
Thus the swimming trajectories can be  characterized by considering the following dynamic system:      $$ \dfrac{dx}{dt}=V_x, \;\dfrac{dz}{dt}=V_z, \;       \dfrac{d\theta_p}{dt}=\Omega_y.$$

In the case of a self-diffusiophoretic micromotor near a wall \citep{Mozaffari2016}, the near-wall adjustments in the concentration distribution provide a cushioning effect against the steric collision between the swimmer and the wall. In contrast, in the present situation, the swimmer often approaches very close to the wall, leading to an inevitable crashing against the wall, and no further information about its motion can be retrieved.
Such a contrast arises due to different physical conditions of the concentration and temperature fields at the plane wall in the respective cases. In the case of a comparable self-diffusiophoretic swimmer, the physical circumstance of the impermeability of solutes at the wall demands a zero flux boundary condition. In contrast, here the wall is maintained isothermal by using a constant temperature bath, therefore allowing a finite normal temperature gradient at the wall.
The dimensionless heat flux absorbed by the cold isothermal wall from the heated fluid takes the form
\begin{equation}\label{eq:wall_flux}
q''_w= \frac{\partial \mathcal{T}_{f}}{\partial z}\bigg{|}_{z=0}= \frac{1-\cos(\eta)}{\sinh(\xi_0)}\frac{\partial T_f}{\partial \xi} \bigg{|}_{\xi=0},
\end{equation}
which upon using \eqref{eq:Temp_out_2} reduces to
\begin{equation}
q''_w=\frac{(1-\cos(\eta))^{3/2}}{2\sinh(\xi_0)} \sum_{m=0}^{\infty}\sum_{n=m}^{\infty} \left(2 \,n+1\right) A_{n,m} P_n^{m}(\cos\eta)\cos(m\phi).
\end{equation}
The resulting differences in the temperature distribution keep the whole wall-adjacent region cool, conforming to the boundary condition of a cold wall.
Such a thermal constraint nullifies the accumulation of heat in the gap between the particle and wall, which would have caused a fluid flow conducive to the near-wall cushioning characteristics.

The problem of steric collision has been circumvented by employing  an electrostatic-type, short-ranged repulsive force at the plane wall \citep{Spagnolie2012} $: \mathbf{F}_\mathrm{rep}= \dfrac{\alpha_1 \exp{(-\alpha_2 \, \delta)}}{1-\exp{(-\alpha_2 \, \delta)}} \mathbf{\hat{e}_z} $.  The parameters  $ \alpha_1,\alpha_2$  have typical values  $200,100$, respectively, so that the swimmer does not approach closer than $ \sim 0.01 $ times the swimmer radius. 
Such a scenario was previously encountered by others in relation to `squirmers'\citep{Spagnolie2012,Li2014,Poddar2020} as well as self-diffusiophoretic microswimmers \citep{Ibrahim2016}, and different forms of repulsive potentials were employed. It is noteworthy that the squirmer models only deal with the hydrodynamics of the microswimmer, and the wall-induced distortion of the scalar field (e.g. temperature or solute) is not captured. Hence, the wall-bound motion of an auto-thermophoretic microswimmer cannot be predicted by either of the self-diffusiophoretic or the squirmer models.

Some of the limiting circumstances in the micromotor trajectory have to be treated cautiously.
As $ \varphi_{cap} \to 180^\circ $ the temperature asymmetry about the director $ \mathbf{{d}} $ vanishes completely, and motion of the particle is similar to a uniformly heated particle near a wall.
In this condition,  asymmetric  distribution of the driving  influences provides an attractive force along the '$-z$' direction, leading to a 'direct impact' onto the wall. Virtually, a similar scenario arises if the micromotor is  launched with its director pointing along $ -z$ (i.e. $ \theta_{p,0}=180^\circ $). The condition $ \theta_{p,0}=0^\circ $ is relatively involved. In that case, consistent with the common intuition, the swimmer escapes along an upward straight line as long as the cap coverage angle ($ \varphi_{cap}$) is below a critical value. Beyond this $\varphi_{cap}$, the downward attraction effect due to the wall (discussed later) becomes prominent, and a downward direct impact occurs. The `direct impact' and `upward escape' swimming states have been denoted by black and blue triangles, respectively in the phase diagrams to follow.
\subsubsection{Swimming state transitions due to thermal conductivity contrast}
\label{ssec:transition_swim_K}
In the following discussion, different swimming trajectories are categorised and described focusing on the results where a characteristic shift of the trajectories takes place due to the variations in the particle to fluid thermal conductivity ratio.

\subsubsection*{$\blacksquare \;$ {Case-I}: From sliding to escape with a small heated cap}
\begin{figure}[!htb]
    \centering
    \begin{subfigure}{0.6\textwidth}
        \centering
        \includegraphics[width=0.75\textwidth]{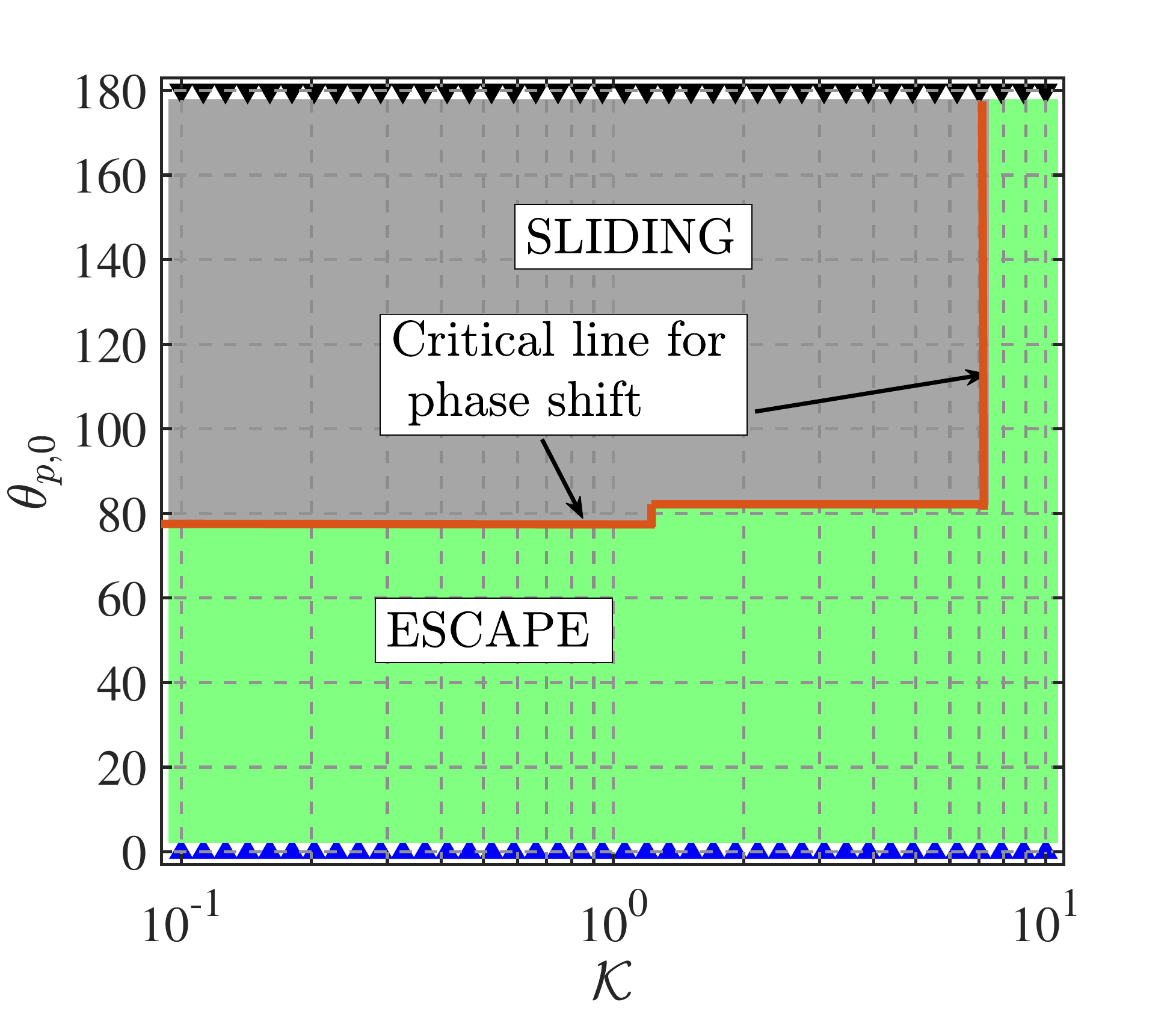}
         \vspace{14ex}
        \caption{ }
        \label{fig:Phase_diag_CASE_1}
    \end{subfigure}
\vspace{2ex}
    \begin{subfigure}{0.55\textwidth}
        \centering
        \includegraphics[width=1\textwidth]{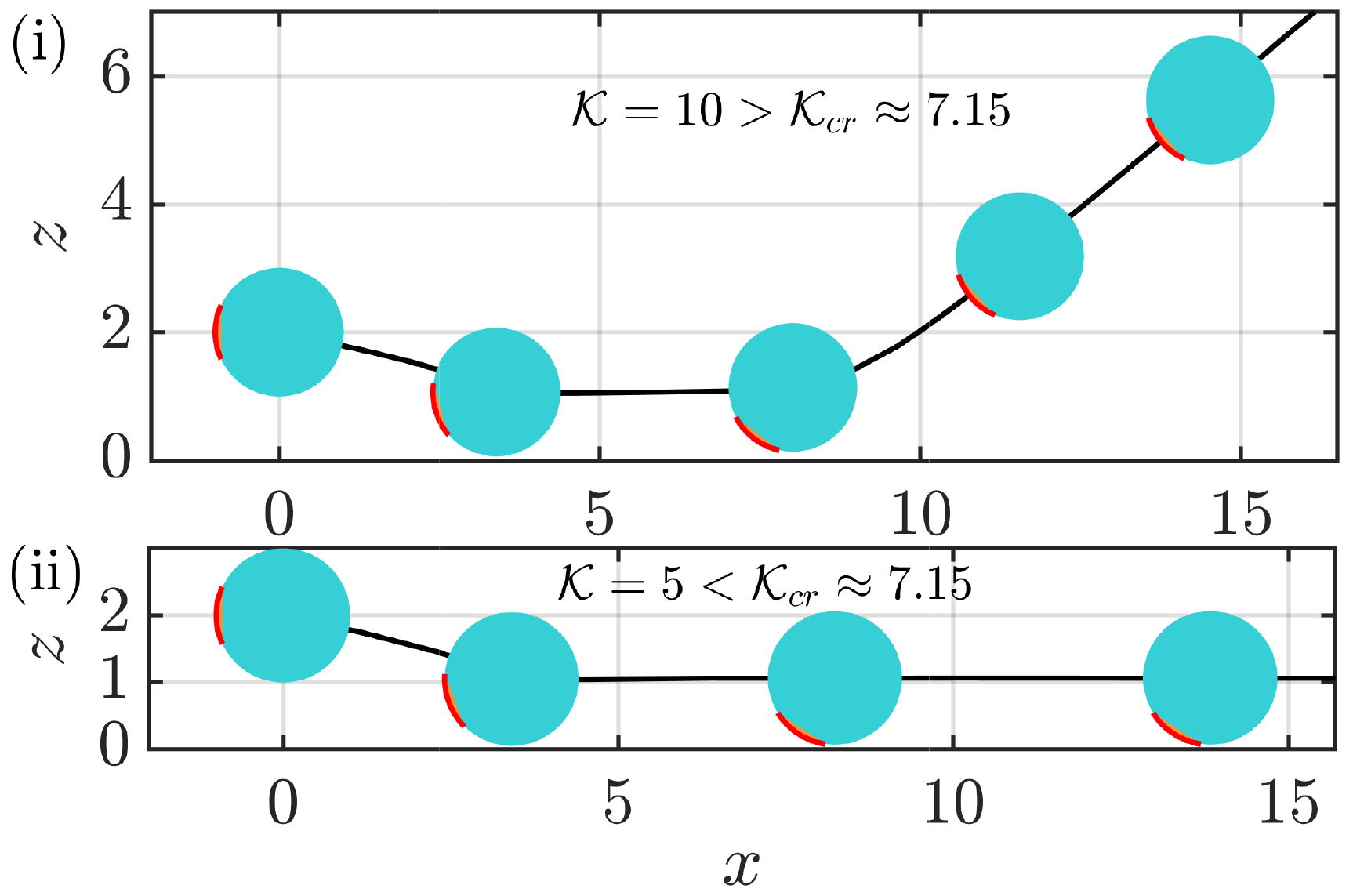}
         \vspace{10ex}
        \caption{ }
        \label{fig:lambda_5_10_slide}
    \end{subfigure}
    \begin{subfigure}{0.43\textwidth}
        \centering
        \includegraphics[width=1\textwidth]{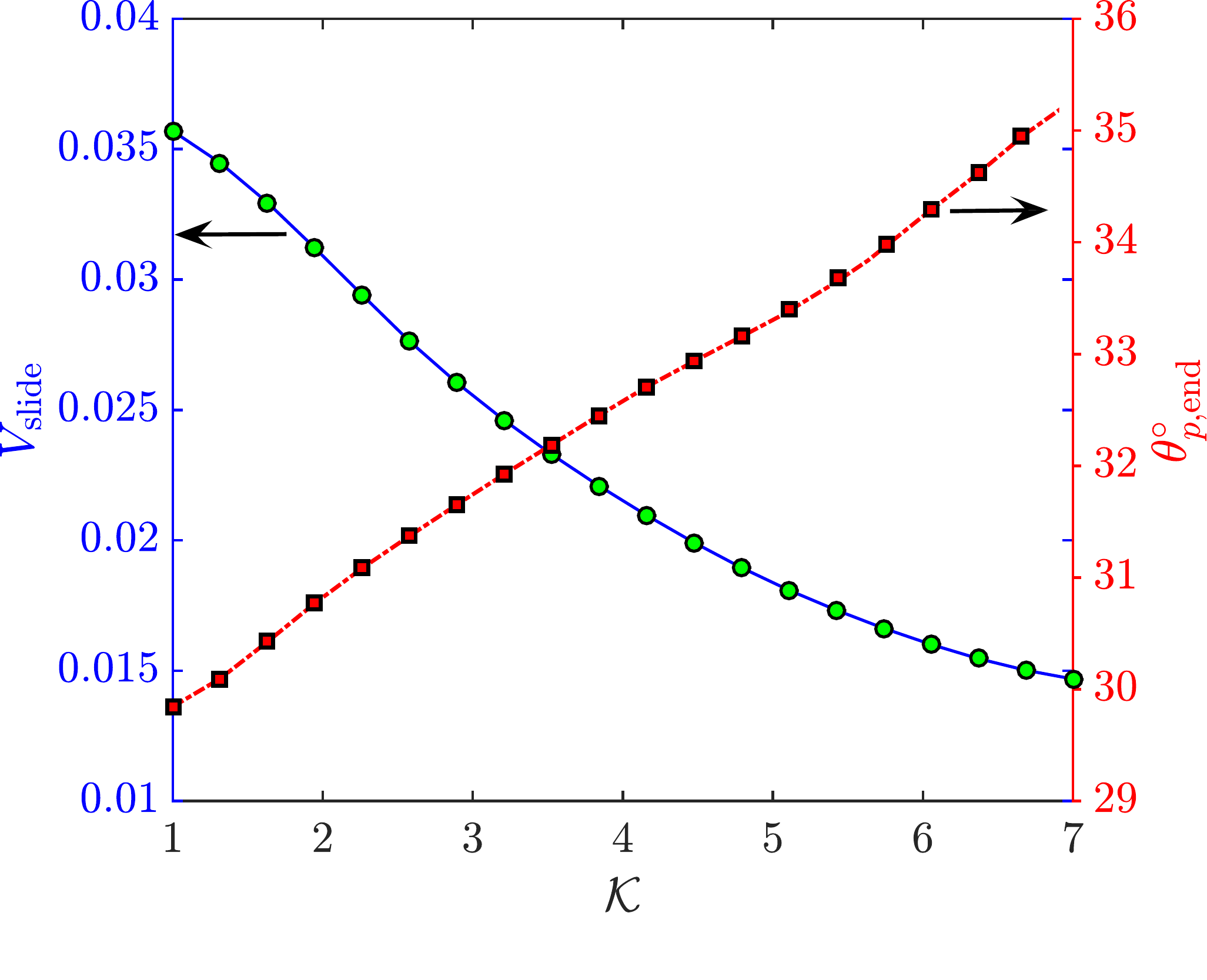}
        \vspace{8ex}
        \caption{}
        \label{fig:Slide_vx_th_Case_1}
    \end{subfigure}
    \vspace{-1ex}
    \caption[CASE-I: (a) Phase map of the microswimmer trajectories on the $ \mathcal{K}-\theta_{p,0} $ plane. (b) Comparison of swimming trajectories for different thermal conductivity contrasts with $ \theta_{p,0}=90 ^{\circ}$.
    (c) Variation of sliding velocity $ (V_\mathrm{slide}) $ and final orientation angle $ (\theta_{p,\mathrm{end}}) $ with $ \mathcal{K} $ for $<\mathcal{K}_{cr}$ as per subfigure (b).]
    {{Case-I}: (a) Phase map of the microswimmer trajectories on the $ \mathcal{K}-\theta_{p,0} $ plane. (b) Comparison of swimming trajectories for different thermal conductivity contrast with $ \theta_{p,0}=90 ^{\circ}$. In panel (i): $ \mathcal{K}=10 $ and (ii): $ \mathcal{K}=5 $. In panel (i) the microswimmer escapes from the wall and attains a fixed orientation $ \theta_p= 48.7 ^{\circ}$. In panel (ii), it slides along the wall with a fixed velocity of $ V_{slide}= 0.018$ with a fixed orientation, $ \theta_{p,end}= 33.23 ^{\circ}$. For both (a) and (b) we have chosen $ \varphi_\text{cap}=25^{\circ} $ and $ h_ 0=2$.
        (c) Variation of sliding velocity $ (V_\mathrm{slide}) $ and final orientation angle $ (\theta_{p,\mathrm{end}}) $ with $ \mathcal{K} $ for $<\mathcal{K}_{cr}$ as per subfigure (b).}
    \label{fig:case_1}
\end{figure}

In figure~\ref{fig:case_1}(a), the swimming trajectories of the micromotor are illustrated with the help of a phase diagram on the $ \mathcal{K}-\theta_{p,0} $ plane for a cap coverage angle of $ \varphi_{cap}=25^\circ$ and initial launching height of $ h_0=2 $. For low values of $ \mathcal{K} $, until a launching orientation $ \theta_{p,0} \lesssim 77.5^\circ $ \footnote{The smallest gap considered is $ 2.5^\circ $ for $ \theta_{p,0}$.}, the micromotor only escapes away from the wall, never to return. However, when the swimmer is launched from an orientation leaning more towards the wall, the swimmer slides along the wall keeping a small gap from it and maintains a fixed angular orientation. The scenario changes as the thermal conductivity ratio $ \mathcal{K} $ increases beyond a critical value of $\mathcal{K}_{cr} \approx 1.30$~\footnote{The critical values of $ \mathcal{K} $ are found with a fine resolution until an accuracy of first two significant digits is obtained.}. With the initial orientations more towards the wall, the $\mathcal{K}_{cr} $ value becomes higher, and eventually for $\theta_{p,0} \gtrsim 82.5 ^\circ$, the sliding to escape transition onsets at $\mathcal{K}_{cr} \approx 7.15 $.

Figure~\ref{fig:case_1}(b) demonstrates two typical trajectories of a micromotor launched from $ \theta_{p,0}=90^\circ $ with $ \mathcal{K} $ values below or above the critical one. In both the cases, the micromotor shows an impending motion with $ \Omega_y<0 $ and $ V_z<0$.
Now, similar to our previous discussions in section~\ref{ssec:temp_result},
{a micromotor with $ \mathcal{K}=5 $ experiences stronger surface temperature gradient in comparison to its $ \mathcal{K}=10 $ counterpart. This condition promotes a subsequent raise in the horizontal component of micromotor velocity $(+V_x)$ as well as causes a faster migration in the vertically downward direction $ (-V_z) $.
    {Under the competitive driving forces for rotation (refer to section~\ref{ssec:vel_result}), once a critical condition is reached, the micromotor experiences a net-zero rotation rate $\Omega_y=0$.} {Here the hydrodynamic resistance to the downward approach is not sufficient to prevent a steric collision, and the repulsive force $ (\mathbf{F}_{rep}) $ plays its role by contributing to the upward resistive force, leading to a net-zero vertical velocity, i. e. $ V_z=0$.}
    In contrast to this, with $\mathcal{K}=10$, a micromotor with a lesser amount of counter-clockwise rotation also experiences a reduced hydrodynamic resistance but continues to rotate while remaining in the wall-adjacent region. Eventually, the angular orientation reaches a point where the particle gains an upward velocity component $(V_z>0)$, and it is eventually reflected from the wall. As it moves upward, the rotation becomes diminishingly small, thereby letting the particle to move with a fixed orientation of $ 48.7^\circ $.
}

The increasing particle conductivity not only causes a characteristic shift of the trajectories but also results in modulations in the sliding velocity and the fixed tilt angle with which the swimmer traverses (see figure~\ref{fig:case_1}(d)). As the particle to fluid thermal conductivity ratio increases towards the critical value, the sliding velocity diminishes, and the director tilts more towards the wall.

\subsubsection*{$\blacksquare \;$ {Case-II}: From stopping to escape with small heated cap}

\begin{figure}[!htb]
    \centering
    \begin{subfigure}{0.45\textwidth}
        \centering
        \includegraphics[width=1\textwidth]{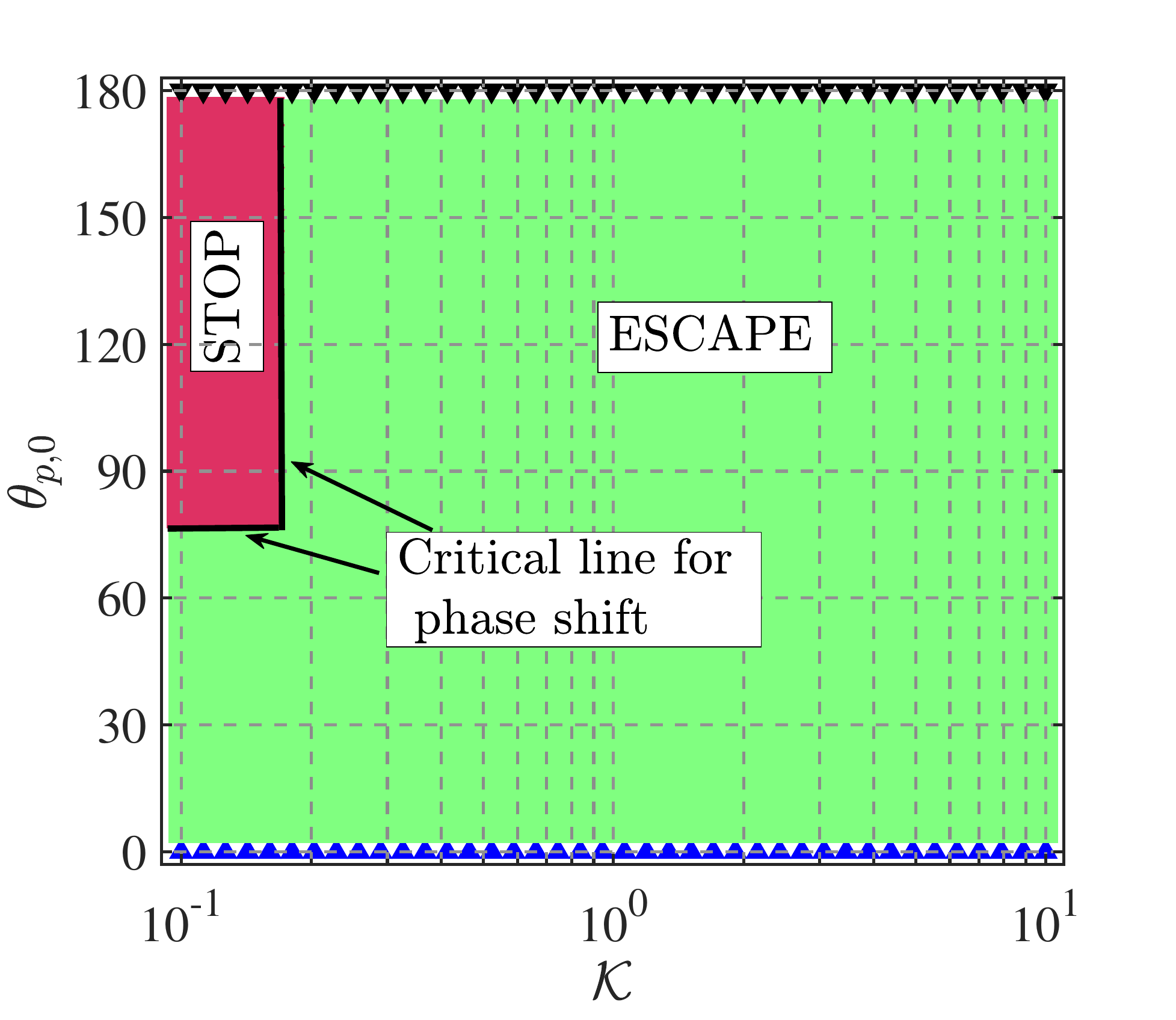}
         \vspace{10ex}
        \caption{}
        \label{fig:Phase_diag_CASE_3}
    \end{subfigure}
    \quad
    \begin{subfigure}{0.45\textwidth}
        \centering
        \includegraphics[width=1\textwidth]{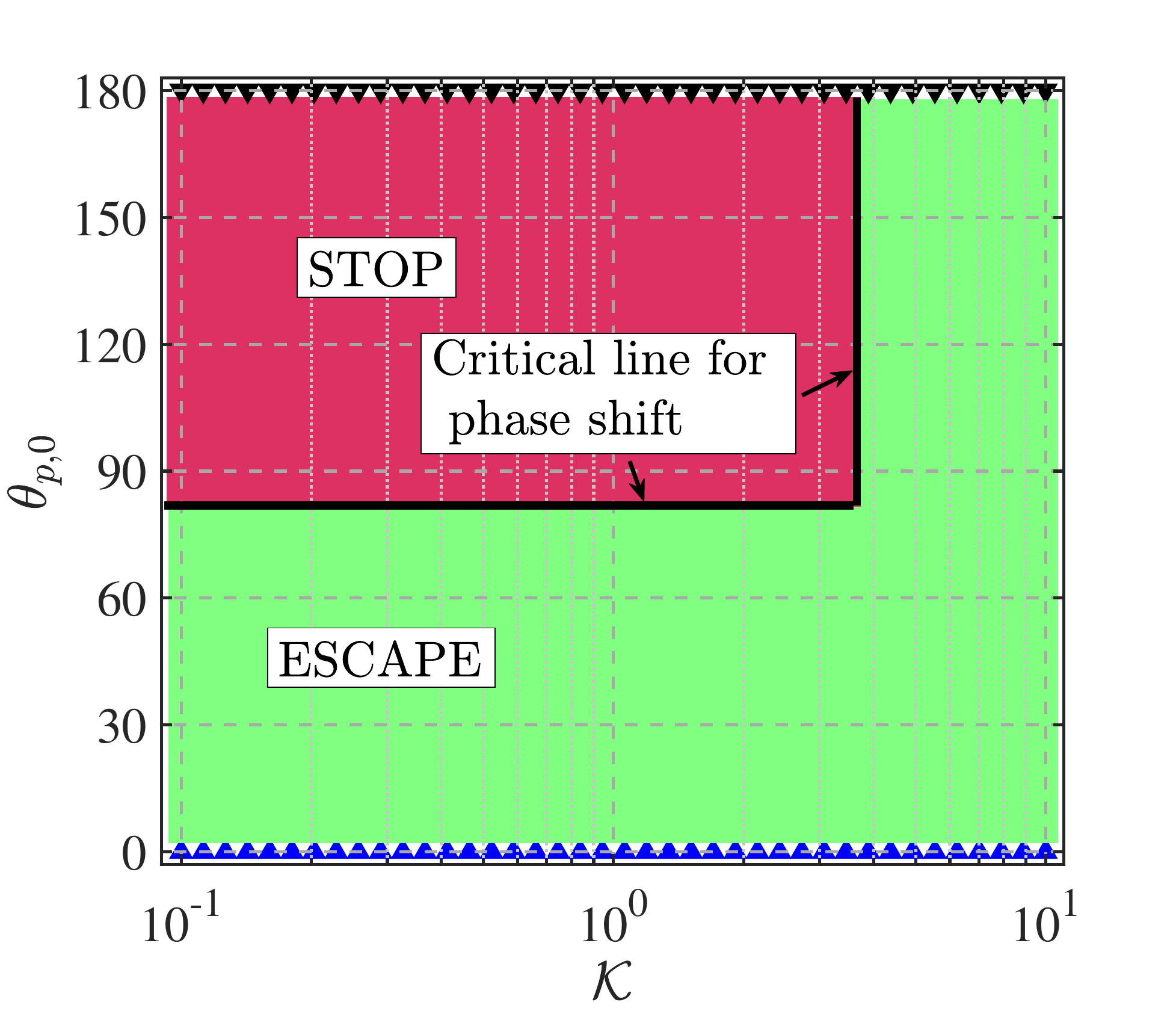}
         \vspace{10ex}
        \caption{}
        \label{fig:Phase_diag_CASE_2}
    \end{subfigure}
    \qquad
    \begin{subfigure}{0.65\textwidth}
        \centering
        \includegraphics[width=1\textwidth]{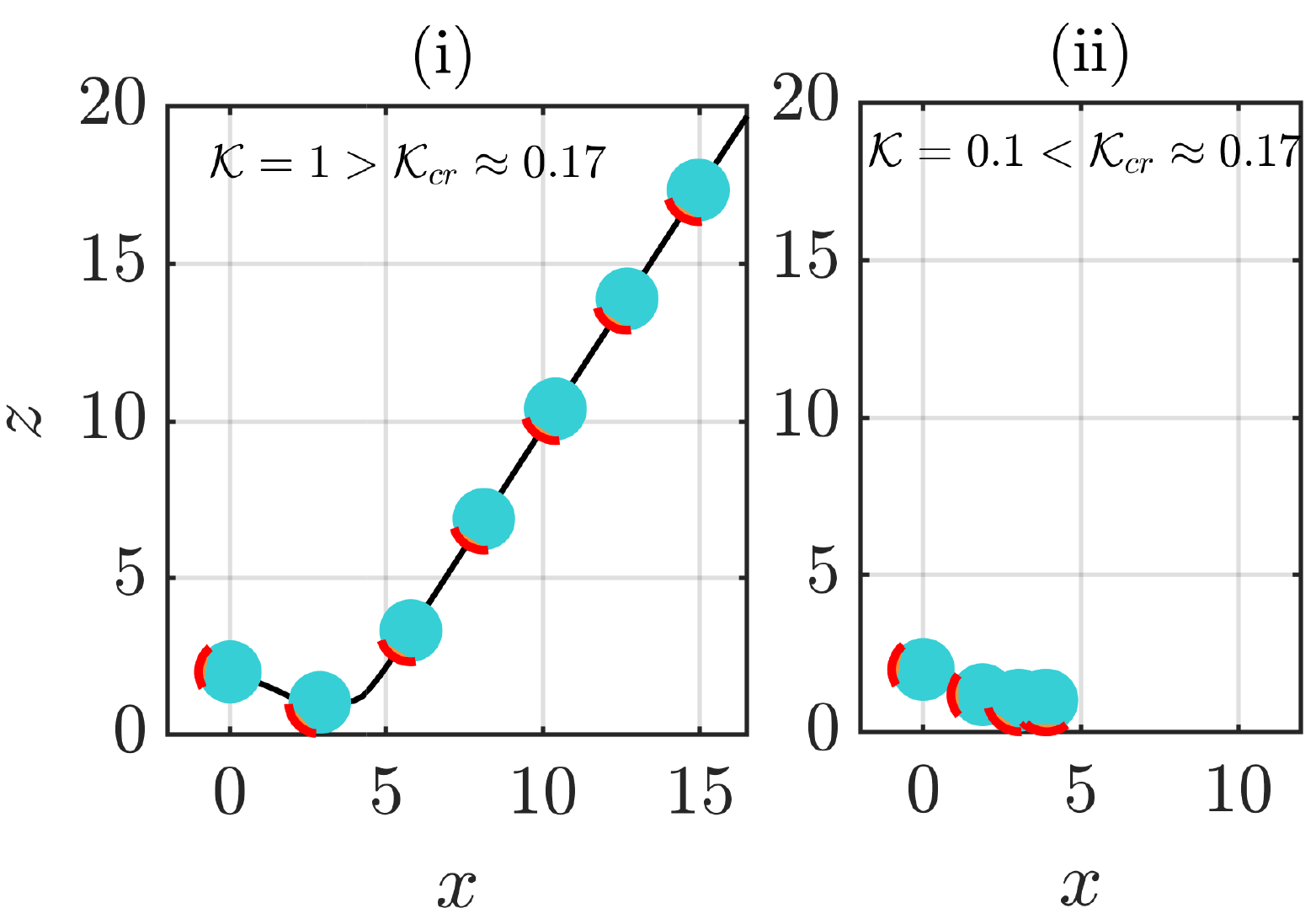}
         \vspace{10ex}
        \caption{}
        \label{fig:Trajectory_lambda_1_0p1_Tp_100_Tcap_40_new}
    \end{subfigure}
    \caption[CASE-II: (a),(b): Phase map of the microswimmer trajectories on the $ \mathcal{K}-\theta_{p,0} $ plane for coverage angles $\varphi_\text{cap}=40 ^{\circ}$ and $50 ^{\circ}$, respectively. (c): Comparison of swimming trajectories for different thermal conductivity contrast with $ \theta_{p,0}=100 ^{\circ}$ and $\varphi_\text{cap}=40 ^{\circ} $.]
    {{Case-II}: (a),(b): Phase map of the microswimmer trajectories on the $ \mathcal{K}-\theta_{p,0} $ plane for coverage angles $\varphi_\text{cap}=40 ^{\circ}$ and $50 ^{\circ}$, respectively. In both the figures and $h_0=2$.
        (c) Comparison of swimming trajectories for different thermal conductivity contrast with $ \theta_{p,0}=100 ^{\circ}$ and $\varphi_\text{cap}=40 ^{\circ} $. In panel (i): $ \mathcal{K}=1 $ and (ii): $ \mathcal{K}=0.1 $. In panel (i) the microswimmer escapes from the wall and attains a fixed orientation $ \theta_p= 36.5 ^{\circ}$. In panel (ii) it becomes stationary at position $ (x_{stop},\delta_{stop})=(3.92,0.027) $ and a vertical fixed orientation, i.e. $ \theta_p \approx 0 ^{\circ}$.}
    \label{fig:case_2}
\end{figure}

\noindent Here, we present the case when the micromotor has a small heated cap, and a transition from stationary to escaping trajectories takes place for a critical value of the thermal conductivity ratio $ \mathcal{K}_{cr} $. This critical condition can be realized for either of the conditions - (A) the fluid is more conductive than the particle (demonstrated in figure~\ref{fig:case_2}(a)) or (B) the particle is more conductive than the fluid (\ref{fig:case_2}(b)). Comparing figures~\ref{fig:case_2}(a) and (b), we find that the critical thermal conductivity contrast shifts from $ \mathcal{K}_{cr}\approx 0.17$ to 3.7, as the heated area increases from $ \varphi_{cap}=40^\circ$ to $ 50^\circ$. Figure~\ref{fig:case_2}(c)-(ii) depicts that, before the critical condition in the phase diagram is reached $ (\mathcal{K}<0.17)$, the micromotor comes very close to the wall while rotating in the counter-clockwise direction under strong thermophoretic propulsive torque. Soon it reaches a condition where the director $ (\mathbf{ d}) $ points vertically upward, i.e. $ \theta_{p,end}=0$. Such an equilibrium orientation lacks any driving force for axial migration (i.e. $ V_x=0$) as well as rotation (i.e. $ \Omega_y=0$). At the same time, the downward phoretic thrust force is balanced by the tremendous hydrodynamic resistive force and the repulsive potential of the wall $ \mathbf{F}_\text{rep}$. Thus, the motion of the swimmer is completely arrested, and it reaches a 'stationary' state. However, when the $ \mathcal{K}$ value increases beyond $ \mathcal{K}_{cr}$ (refer to Figure~\ref{fig:case_2}(c)-(ii)), the counter-clockwise rotation is weaker, and it migrates downward at a slower pace. As a result, well before the orientation reaches the $ \theta_p=0$ state, the micromotor attains the critical position for the reversal of $-V_z$ to $+V_z$, a phenomenon causing a reflecting trajectory similar to figure~\ref{fig:case_1}(b)-(i).

\subsubsection*{$\blacksquare \;$ {Case-III}: From stopping to escape with a large heated cap}
\begin{figure}[!htb]
    \centering
    \begin{subfigure}{0.45\textwidth}
        \centering
        \includegraphics[width=1\textwidth]{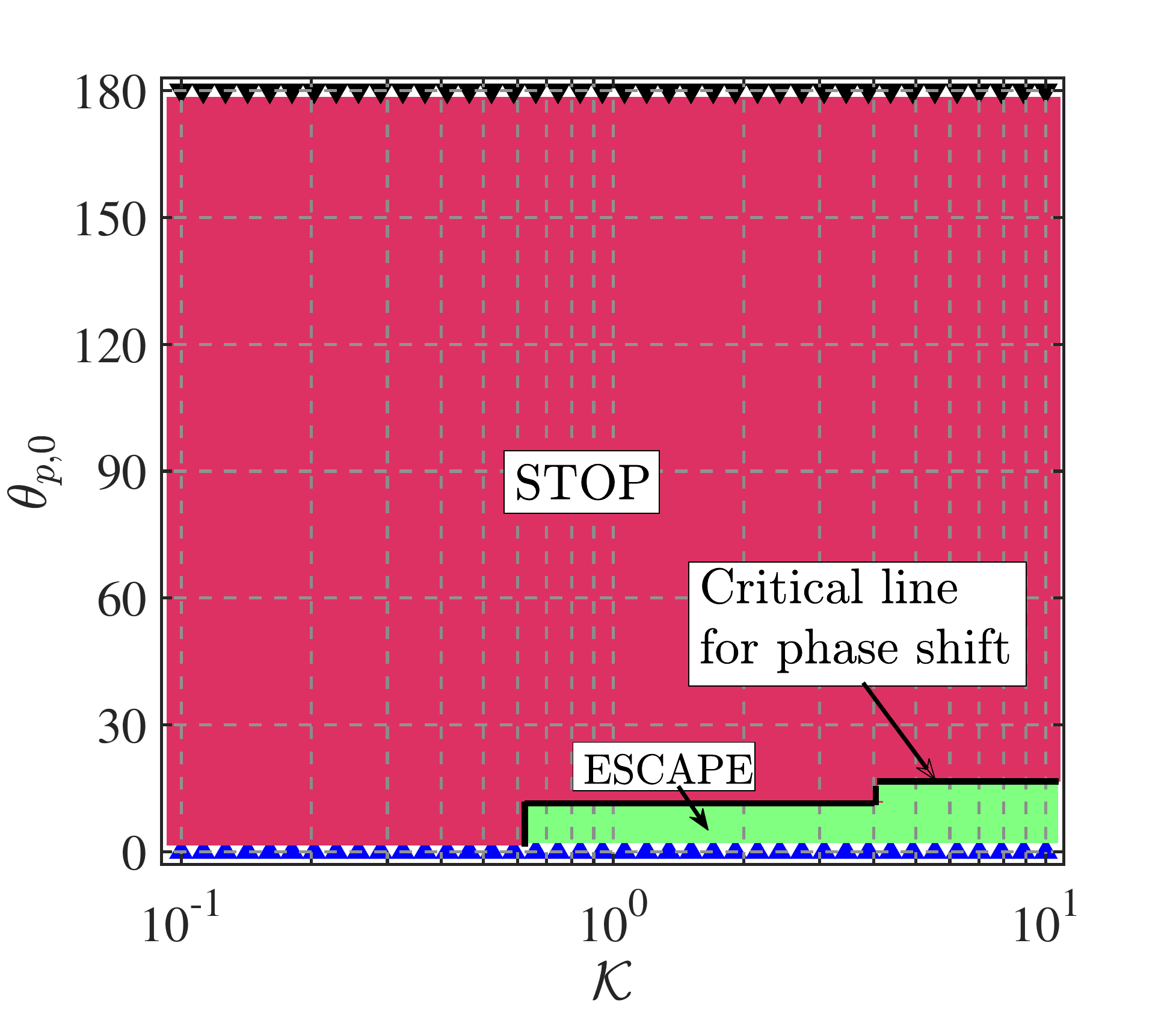}
        \vspace{7ex}
        \caption{}
        \label{fig:Phase_diag_CASE_4}
    \end{subfigure}
    \qquad
    \begin{subfigure}{0.45\textwidth}
        \centering
        \includegraphics[width=1.3\textwidth]{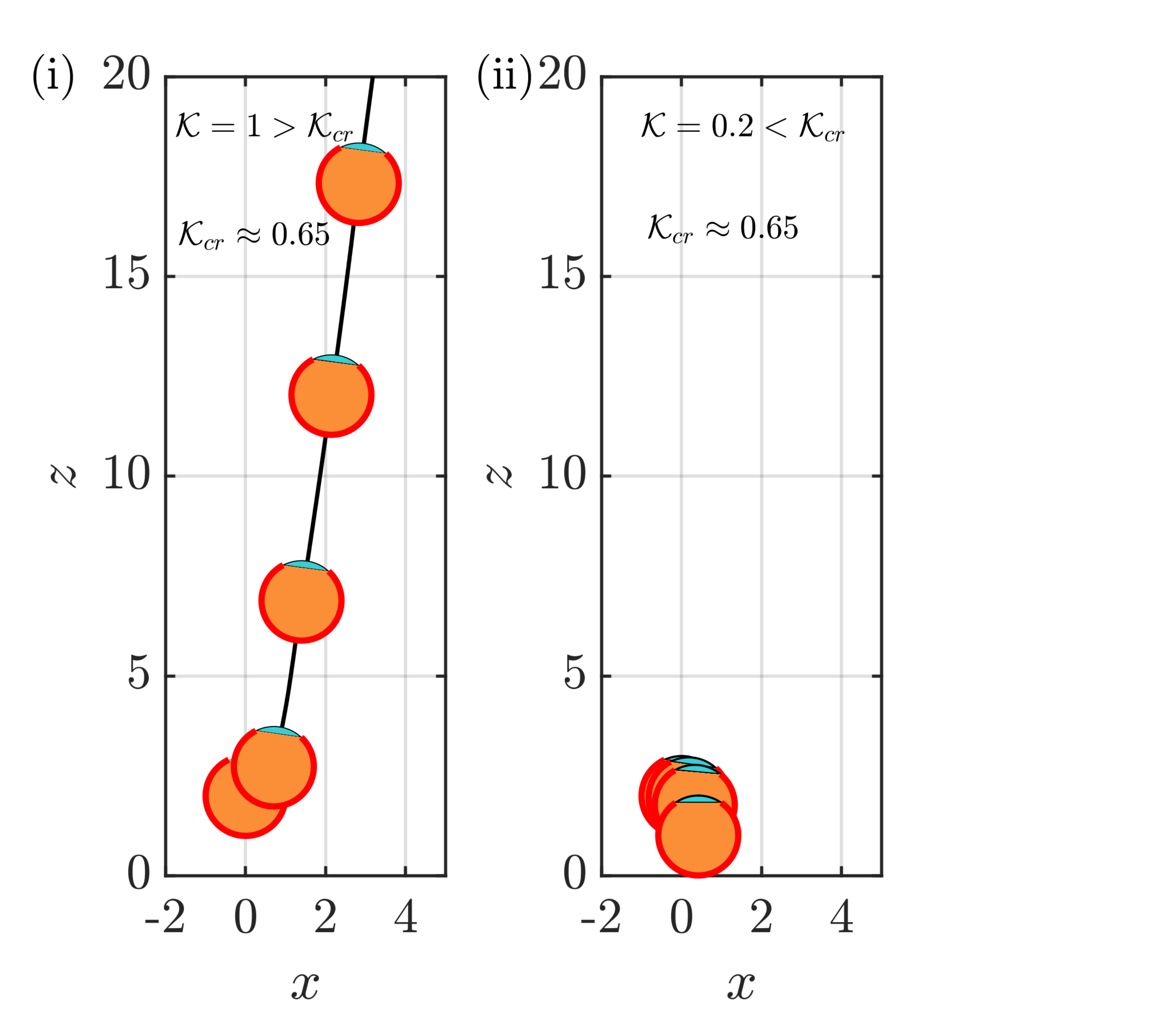}
         \vspace{6ex}
        \caption{}
        \label{fig:Trajectory_lambda_1_0p2_Tp_10_Tcap_145}
    \end{subfigure}
    \vspace{-1ex}
    \caption[CASE-III: (a): Phase map of the microswimmer trajectories on the $ \mathcal{K}-\theta_{p,0} $ plane for coverage angles $\varphi_\text{cap}=145^{\circ}$. (b): Comparison of swimming trajectories for different thermal conductivity contrast with $ \theta_{p,0}=10 ^{\circ},\varphi_\text{cap}=145 ^{\circ} $ and $ h_0=2 $. ]
    {{Case-III}: (a) Phase map of the microswimmer trajectories on the $ \mathcal{K}-\theta_{p,0} $ plane for coverage angles $\varphi_\text{cap}=145^{\circ}$. (b) Comparison of swimming trajectories for different thermal conductivity contrast with $ \theta_{p,0}=10 ^{\circ},\varphi_\text{cap}=145 ^{\circ} $ and $ h_0=2 $. In panel (i): $ \mathcal{K}=1 $ and (ii): $ \mathcal{K}=0.1 $. In panel (i), the microswimmer escapes from the wall and attains a fixed orientation $ \theta_p= 7.8 ^{\circ}$. In panel (ii), it becomes stationary at position $ (x_{stop},\delta_{stop})=(0.42,0.015) $ and a vertical fixed orientation, i.e. $ \theta_{p,stop} = 0 ^{\circ}$.}
    \label{fig:case_4}
\end{figure}
\noindent
Here the cap coverage is high $(\varphi_{cap}=145^\circ)$, but the nature of transition is almost similar to the {Case-II} presented above. The only distinguishing factor in the transition is the height above the wall $ (\delta_{cr}) $, where the sign flip of $V_z$ takes place for $ \mathcal{K}>\mathcal{K}_{cr}$. Similar to {Case-I}, $\mathcal{K}_{cr}$ is a function of the initial orientation $ (\theta_{p,0})$. Figure~\ref{fig:Phase_diag_CASE_4} shows that, in the domain $ 0<\theta_{p,0}<12.5^\circ$, we have $ \mathcal{K}_{cr} \approx 0.65$, while for $ 12.5^\circ \lesssim \theta_{p,0}<17.5^\circ$, we have $ \mathcal{K}_{cr} \approx 1$. As illustrated in \ref{fig:Trajectory_lambda_1_0p2_Tp_10_Tcap_145}-(i), the micromotor never comes too close to the wall, and the critical height is just below the initial launching height $ (h_0) $. This provides a sense of repulsive action from the wall at a much greater vertical distance than that was in {Case-II}.

\subsubsection{Influence of initial height, cap coverage, and initial orientation}
\label{sssec:height}
\begin{figure}[!htb]
    \centering
    \includegraphics[width=1.1\textwidth]{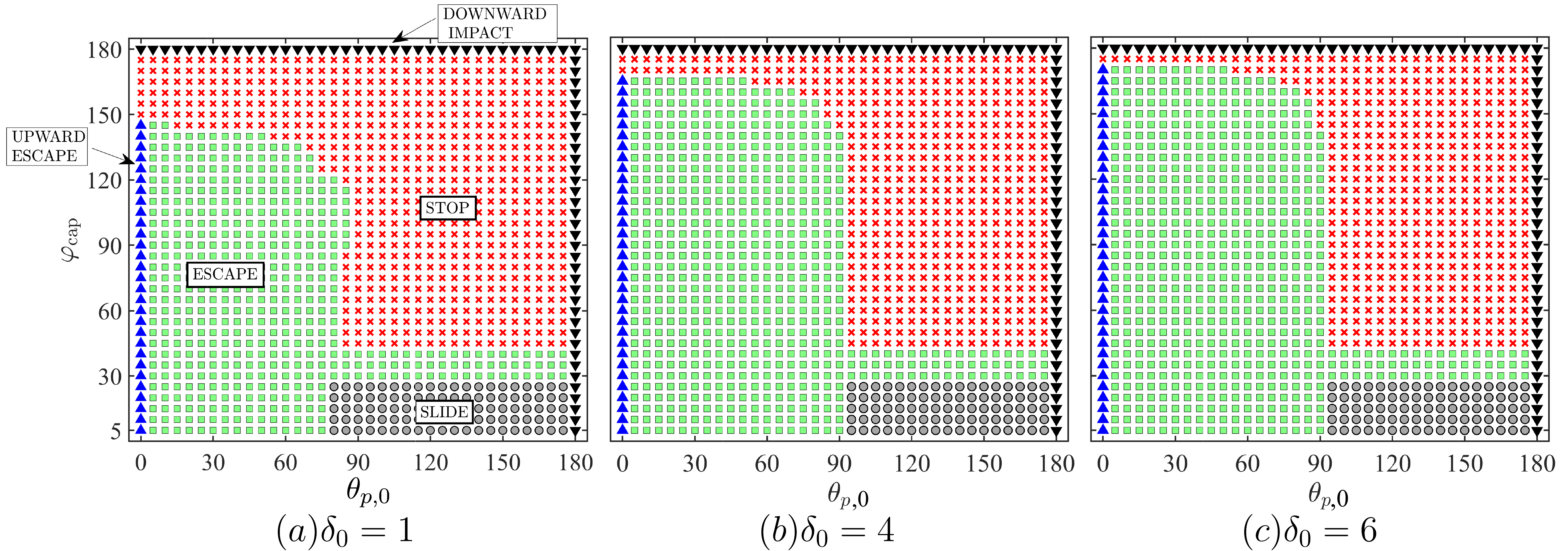}
    \caption[Phase maps categorising the swimming behaviours of the auto-thermophoretic microswimmer for different values of the coverage angle of the metal cap $ (\varphi_\text{cap}) $ and initial launching angle $ (\theta_p) $.]
    {Phase maps categorising the swimming behaviours of the auto-thermophoretic microswimmer for different values of the coverage angle of the metal cap $ (\varphi_\text{cap}) $ and initial launching angle $ (\theta_p) $. Panels (a),(b) and (c) correspond to different initial heights as shown. The thermal conductivity ratio has been chosen as $ \mathcal{K}=1$. Here the phase maps have been constructed on a grid of $ 5^\circ $ intervals on both the coordinates.}
    \label{fig:Phase_diag_lambda_146}
\end{figure}
\begin{figure}[!htb]
    \centering
    \includegraphics[width=0.75\textwidth]{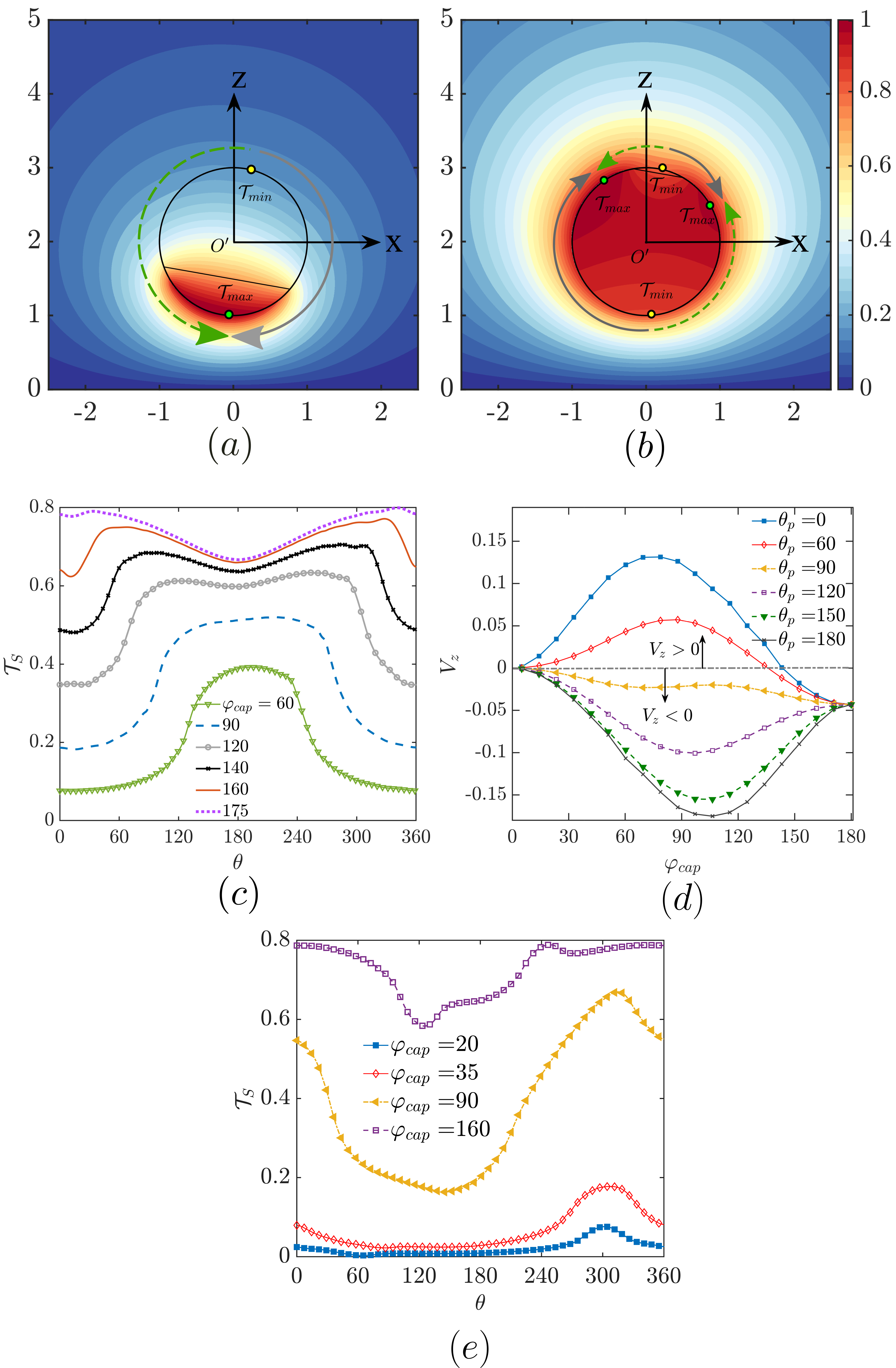}
    \caption[(a), (b): Showing the direction of surface flow for low and high coverage angles $ \varphi_{cap}=60^\circ $ and $160^\circ $, respectively.
    (c): Variation of surface temperature along the surface of the micromotor for different coverage angles $ (\varphi_{cap}) $. (d): Variation of the vertical component of the micromotor velocity $ (V_z) $ with $ \varphi_{cap} $ for different $ \theta_{p}$. (e): Variation of surface temperature along the surface of the micromotor for different coverage angles $ (\varphi_{cap}) $ with $ \theta_{p}=120^\circ$. ]
    { (a), (b): Showing the direction of surface flow for low and high coverage angles $ \varphi_{cap}=60^\circ $ and $160^\circ $, respectively. Here the background colour denotes the scaled temperature field.
        (c): Variation of surface temperature along the surface of the micromotor for different coverage angles $ (\varphi_{cap}) $. $ \theta_{p} =10 ^ \circ$ has been chosen for (a)-(c). (d): Variation of the vertical component of the micromotor velocity $ (V_z) $ with $ \varphi_{cap} $ for different $ \theta_{p}$. (e): Variation of surface temperature along the surface of the micromotor for different coverage angles $ (\varphi_{cap}) $ with $ \theta_{p}=120^\circ$. Other parameters are chosen as $ \delta=1 $ and $ \mathcal{K} =1$.}
    \label{fig:COMPACT_ts_compare_tcap_vary}
\end{figure}
The final swimming feature of the micromotor has been summarized in figures~\ref{fig:Phase_diag_lambda_146}(a)-(c) for all the possible ranges of the heated cap angle $\varphi_\textrm{cap}$ and the initial launching angle $\theta_{p,0}$.
These figures help elucidating some key differences in the trajectory characteristics with those presented previously in connection with the self-diffusiophoresis phenomenon near a solute-impenetrable plane wall \citep{Mozaffari2016,Uspal2015a,Ibrahim2016}.
First, the thermophoretic force is of a strikingly distinct nature in imparting a downward motion to the micromotor even when the director initially faces away from the wall (e.g. in figure~\ref{fig:Phase_diag_lambda_146} (a)).
In the corresponding problems addressed, the micromotor never comes downward if the initial orientation is away from the wall, i.e. $ \theta_{p,0}<90^\circ$. It was also found that the micromotor does not achieve a velocity of $ V_z<0 $ for any extent of the active area if $ \theta_{p,0}<90^\circ$. In sharp contrast to these observations, in the preset scenario, we observe that the micromotor can have $ V_z<0 $ even when $ \theta_{p,0}<90^\circ$. An example of this outcome is provided in figure~\ref{fig:Vz_vs_delta_vary_tp_tcap_90_k_1} for a typical heated cap coverage of $ \varphi_{cap}=90^\circ$ and equal conductivities of the two mediums $ (\mathcal{K}=1)$.
Further, the phase maps in figure~\ref{fig:Phase_diag_lambda_146} (a) demonstrate that, when the micromotor has a large heated cap, the zone of stopping points on the maps penetrates into the region of $ \theta_{p,0}<90^\circ $, and even a fully vertical orientation with $ \theta_{p,0}=0^\circ $ can result in the case of complete stationary trajectories.

As the capped area becomes larger, the distortion of the temperature field becomes intrinsically different, which can be visualized by comparing the temperature profiles in figure~\ref{fig:COMPACT_ts_compare_tcap_vary} (a) and (b).
This is also reflected in the appearance of two distinct maxima points on the $ \mathcal{T}_S \; \text{vs.} \; \theta $ curves for high $ \varphi_{cap} $ instead of just one for low cap angle (please refer to figure~\ref{fig:COMPACT_ts_compare_tcap_vary} (c)). Consequently, the pattern of surface slip flow gets modified as portrayed schematically in figures~\ref{fig:COMPACT_ts_compare_tcap_vary} (a) and (b).
In the former case, the surface slip flow takes place from the colder region (point $ \mathcal{T}_{min}$ in figure~\ref{fig:COMPACT_ts_compare_tcap_vary}(a)) to the hotter region (point $ \mathcal{T}_{max}$ in figure~\ref{fig:COMPACT_ts_compare_tcap_vary}(a)), and the particle gains an upward velocity component $ (V_z>0) $. In contrast to this, for $ \varphi_{cap}=160^\circ $, the continuity of flow demands a pair of new circulating rolls due to a reorganization of the surface temperature distribution.
As a cumulative effect, the downward surface traction surpasses the upward one, and subsequently the motion of the micromotor becomes downward ($ V_z<0$). An involved relationship of $ V_z $ on both $ \varphi_{cap} $ and $ \theta_{p}$ has been presented in figure~\ref{fig:COMPACT_ts_compare_tcap_vary}(d). While this only provides an insight into the impending downward or upward motion during launching, a sliding to escape or escape to stopping transition of swimming states takes place due to qualitatively similar reasons of alterations in particle rotation $ (\Omega_y) $ and vertical motion $ (V_z) $ at a critical transition, as discussed in {Case-I} of section~\ref{ssec:transition_swim_K}.

The mechanism by which changing cap angle $ (\varphi_{cap}) $ influences the swimmer surface flow is more intricate than the corresponding implication due to thermal conductivity contrast $ (\mathcal{K}) $. We have earlier noted that modulations in $ \mathcal{K} $ hold  the capacity of changing only the relative magnitudes of the peak surface temperatures $ (\mathcal{T}_{max}-\mathcal{T}_{min} ) $, but not the peak locations on the swimmer surface (refer to figure~\ref{fig:temp_compact_line}). Conversely, the cap angle $ (\varphi_{cap}) $ variation has the potential of shifting the locations of the peak temperatures. Also, it can create or destroy new peak points (see figure~\ref{fig:COMPACT_ts_compare_tcap_vary}(e) for example), thereby regulating the surface flow to a great extent.

Secondly, during the swimming states where the micromotor ends up being motionless, we observe that the director is pointed vertically downward (i.e. $ \theta_{p,end} = 0^\circ$) in the final state (refer to figures~\ref{fig:Trajectory_lambda_1_0p1_Tp_100_Tcap_40_new}(ii) and \ref{fig:case_4}(ii)). Contrarily,  in the previously reported studies \citep{Mozaffari2016}, the final orientation of the stationary states was found to be $ \theta_{p,end} = 180^\circ$.
Also in the present scenario, the swimmer becomes stationary only after sliding along the wall until it reaches the $ \theta_p=0^\circ$ condition, while remaining at a close distance from the wall ($\delta_{stop} \lesssim 0.1$). However, in the self diffusiophoresis problems, the micromotor reaches `stationary' states with the final separation height being as high as $ \sim 5$ times the particle radius.
The difference between the two categories of problems also stems from inherently distinguished wall-induced modulations in the peak surface temperatures of the micromotor. {Figures~\ref{fig:tcap_var_tmax_loc}(a) and (b) suggest that the minimum temperature is shifted towards the wall and attains almost a $ \theta \approx 180^\circ $ position for both the high coverage angles.} 
Moreover, in stark contrast to the study of \citet{Mozaffari2016}, the maximum-location shifts towards the north pole with an increase in the coverage angle from $\varphi_{cap}=120^\circ$ to $ 150^\circ$. 
Following the directions of arrows and their altered lengths in  figure~\ref{fig:tcap_var_tmax_loc}(b) to (a), we find that the contribution of counter-clockwise surface traction intensifies. This further causes a net negative rotation leading to $ \theta_{p,end}=0^\circ$, and the same can be verified from \ref{fig:Vx_Wy_tcap_150_K_1}(b). Another difference with their work arises from the decreasing values of $ V_x$, as the swimmer approaches the wall (see \ref{fig:Vx_Wy_tcap_150_K_1}(a)).
Subsequently, the micromotor traverses a little distance before coming to rest.

When the micromotor is launched from a sufficient height above the planar wall $(h_0 \sim 10)$, the wall-induced temperature distortions can hardly influence the particle velocity. Accordingly, the micromotor feels a weak attractive force from the wall.
Along similar lines, a comparison of the phase maps in figures~\ref{fig:Phase_diag_lambda_146}(a)-(c) reveals that the zone of stopping trajectories in the common region of $ \theta_p<90^\circ$ and high $ \varphi_{cap}$ shrinks as the initial height gets increased.

\FloatBarrier
\section{Conclusions and remarks}
\label{sec:conclusion}
We have explored the capability of navigating a thermally asymmetric micromotor by a nearby isothermal plane wall. The wall-induced temperature distortion in and around the micromotor causes several distinctive characteristics of motion ranging from sliding along the wall, reaching a stationary configuration to getting repelled by the wall at a fixed orientation.

For high thermal conductivity contrast between the particle and the fluid, an increasingly significant portion the particle senses the effect of localised heating due to laser irradiation. 
This effect, in turn, weakens the temperature gradient and the subsequent phoretic thrust experienced by the micromotor. Our analysis reveals that for a fixed extent of metallic coating and the initial orientation of the micromotor, its final trajectory can be characteristically altered only by changing the particle to fluid thermal conductivity contrast. The critical thermal conductivity contrasts responsible for a shift either from sliding to escape or from stopping to escape states have also been reported. It was found that these critical conditions exist for both the situations of particle becoming more conductive than the fluid or vice versa. Moreover, for some specific heated cap coverages, the critical thermal conductivity contrast depends on the initial tilt of the micromotor. 
 
When a large extent of micromotor surface is coated with the metallic cap, the wall-bound attraction is so strong that migration towards the wall happens, despite the director initially pointing away from the wall. This observation is highly non-intuitive on the backdrop of the previous studies related to self-diffusiophoresis near a wall. 
Moreover, the zone of sliding states in the phase diagram has shifted from a $115^\circ< \varphi_{cap}<150^\circ $ \citep{Mozaffari2016} condition to 
 $0^\circ <  \varphi_{cap} \lesssim 25^\circ $. 
  Also, a window of intermediate $ \varphi_{cap} $ values appears: $ 25^\circ<\varphi_{cap}<40^\circ$, which is sandwiched between the sliding and stopping states.  
  When the micromotor ends up being stationary, it reaches a configuration with the heated surface facing towards the wall.

\textcolor{black}{Since the adiabatic wall effects on the locomotion of a  self-thermophoretic micromotor  may be perceived by drawing analogies with the reported studies on self-diffusiophoretic transport, and the corresponding effects of an isothermal wall are addressed in this study, the present study may be extended to a generic theoretical platform  accommodating arbitrary variations in wall temperature, considering that a generic boundary condition may be represented by a combination of successive isothermal and adiabatic states. 
}

  The fundamental know-how of the temperature distribution in and around a self-thermophoretic micromotor and its trajectory characteristics in the vicinity of a plane obstacle with a known thermal condition may be extremely beneficial in designing a novel microscale thermal sensor. Moreover, the ability to cause critical transitions in the locomotion behaviour based on related thermal and configurational parameters, as explored here, may be exploited to achieve precise control over the navigation of micromotor-based systems for a variety of applications where interaction with a confining boundary is inevitable. Further tuning of these interactions with the aid of thermal noise, topographical alterations \citep{Simmchen2016} and patterning of the thermal boundary conditions at the wall may be harnessed in the future to achieve  desired transitions in swimming states over reduced spatio-temporal scales.

 \section*{Acknowledgement}
	S.C. acknowledges Department of Science and Technology, Government of India, for Sir J. C. Bose National Fellowship.
	\section*{Conflicts of interest}
	The authors declare no conflicts of interest.

\FloatBarrier
\begin{appendices}
    \renewcommand{\thesection}{Appendix}
    \renewcommand\thefigure{A-\arabic{figure}}    
    \setcounter{figure}{0} 
    \renewcommand{\theequation}{A-\arabic{equation}}
    \setcounter{equation}{0}  
\section{Detailed variations of different quantities in the main text}
\begin{figure}[!htb]
    \centering
    \includegraphics[width=0.5\textwidth]{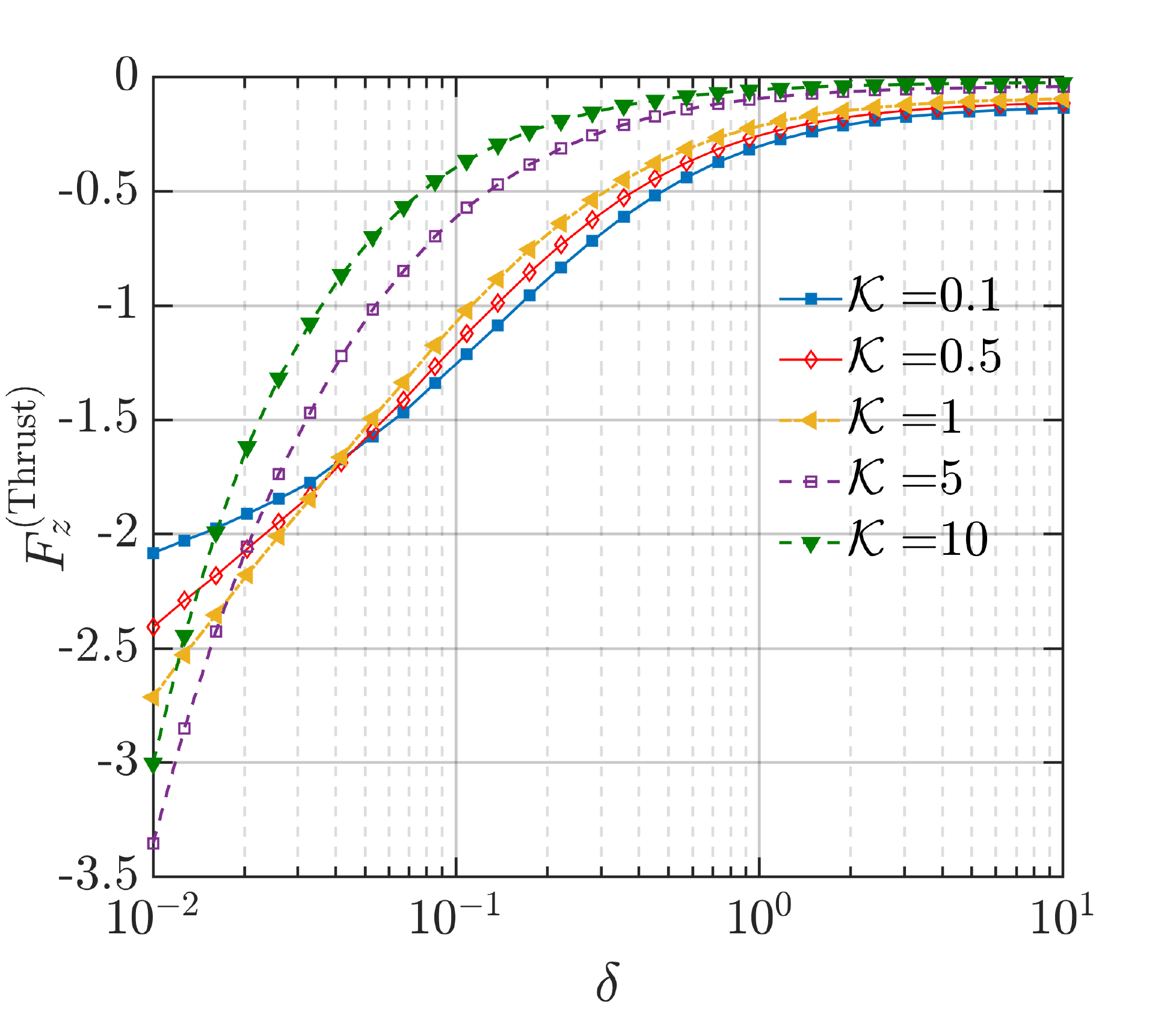}
    \caption[Variation of the vertical thermophoretic thrust $ (F_z^\mathrm{(Thrust)}) $ with $ \delta $ and for different values of $ \mathcal{K} $.]
    {Variation of the vertical thermophoretic thrust $ (F_z^\mathrm{(Thrust)}) $ with $ \delta $ and for different values of $ \mathcal{K} $. The parameters are similar to figure~\ref{fig:Vz_vary_delta_vary_K_tc_90_tp_120}.} 
    \label{fig:Fz_thrust_vary_K_tp_120_tc_90}
\end{figure}
\begin{figure}[!htb]
    \centering
    \includegraphics[width=0.55\textwidth]{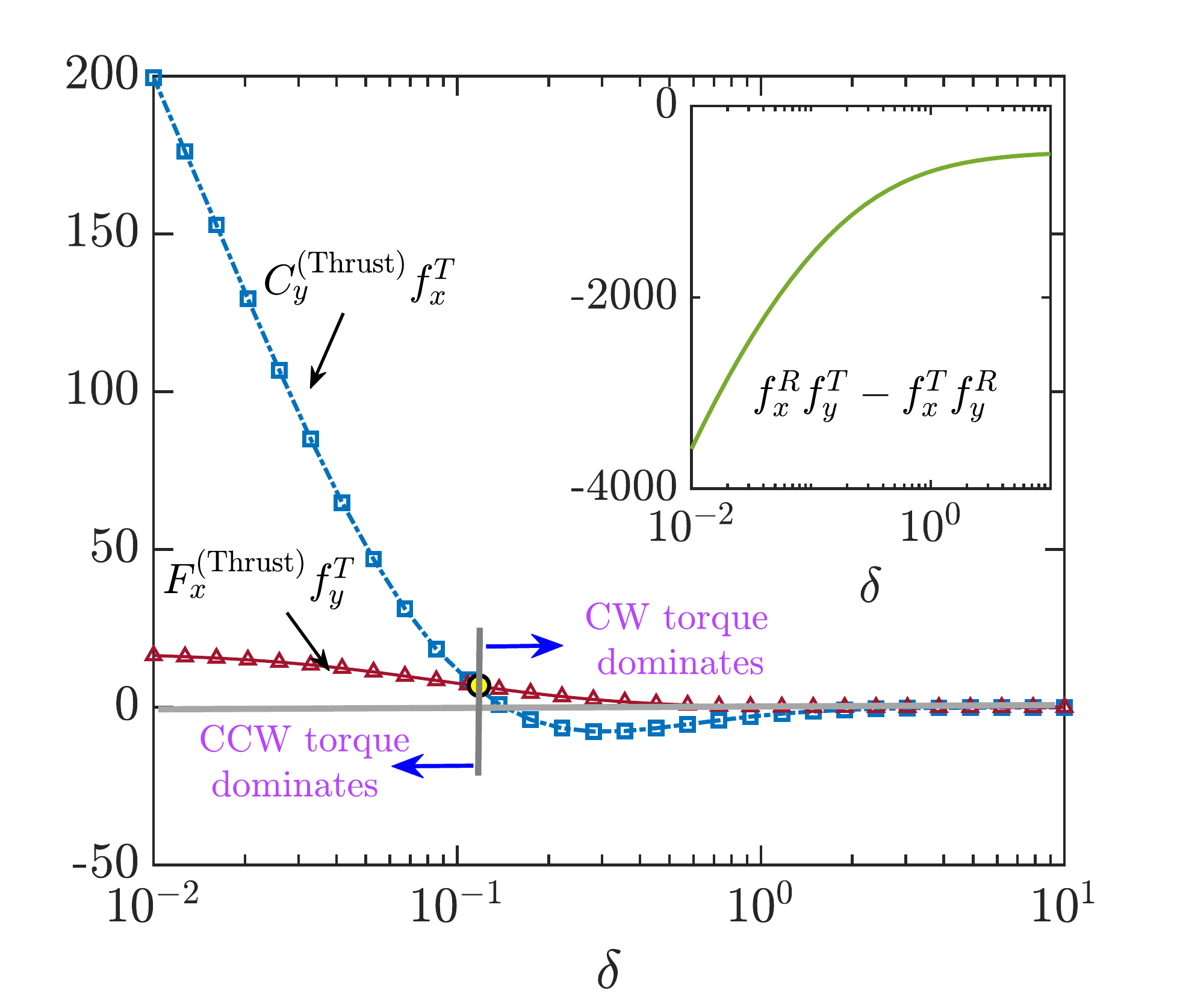}
    \caption[Detailed variation of different terms in \eqref{eq:Vx_Wy_coupled_temp}(b).]
    {Detailed variation of different terms in \eqref{eq:Vx_Wy_coupled_temp}(b). The parameters are similar to figure~\ref{fig:Wy_vary_delta_vary_K_tc_140_tp_150}.  The yellow bubble signifies the location where $C_y^\mathrm{(Thrust)}\times f_{x}^T$ becomes smaller than $F_x^\mathrm{(Thrust)}\times f_{y}^T$.} 
    \label{fig:terms_torque_tp_150_tc_140_K_0p1}
\end{figure}
\begin{figure}[!htb]
    \centering
    \includegraphics[width=0.55\textwidth]{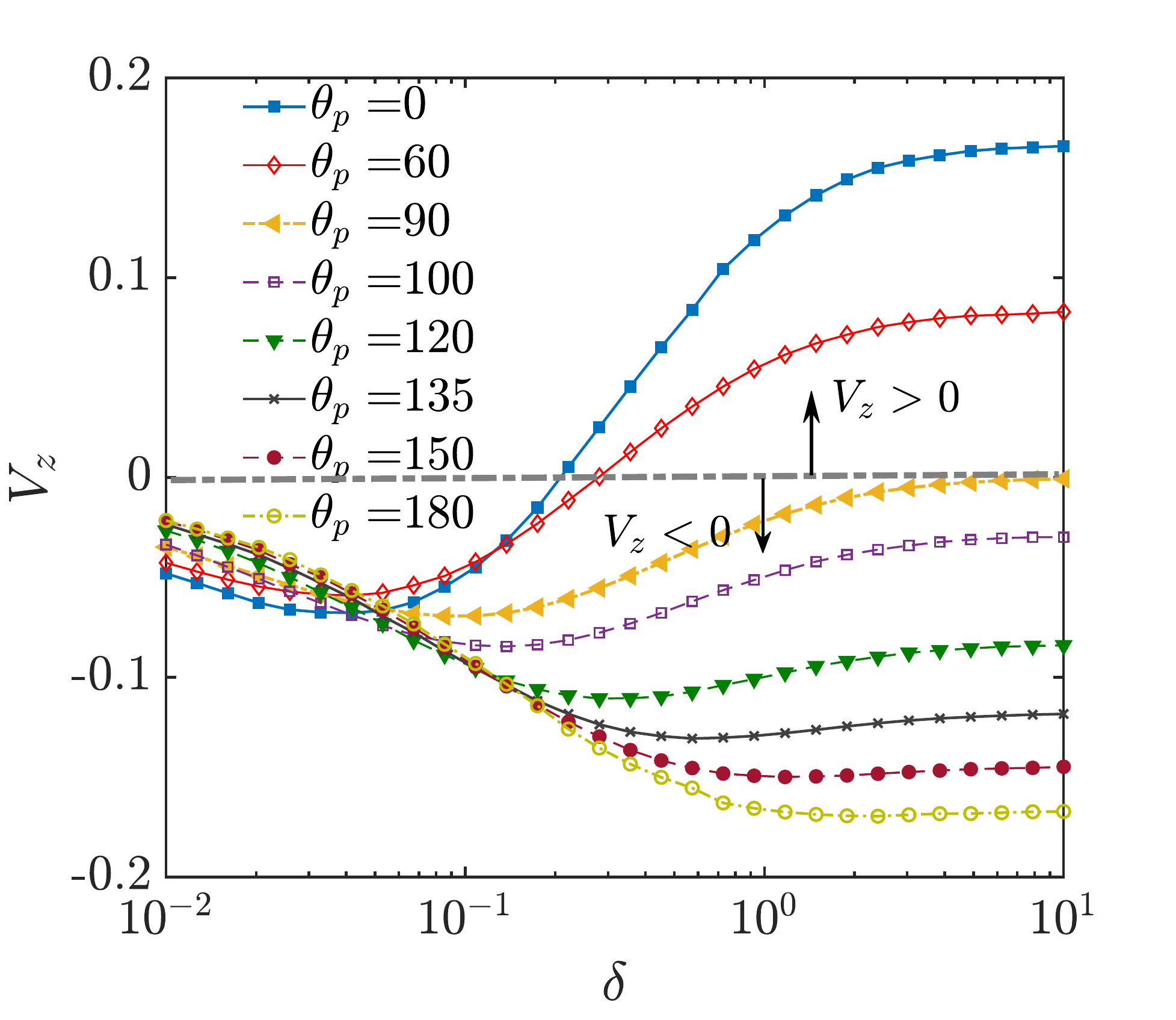}
    \vspace{+2ex}
    \caption{Vertical microswimmer velocity vs. its distance from the wall for different orientation angle $ (\theta_p)$. Here the fixed parameters are $ \varphi_{cap}=90^\circ$ and $ \mathcal{K}=1.$} 
    \label{fig:Vz_vs_delta_vary_tp_tcap_90_k_1}
\end{figure}
\begin{figure}[!htb]
    \centering
    \includegraphics[width=0.75\textwidth]{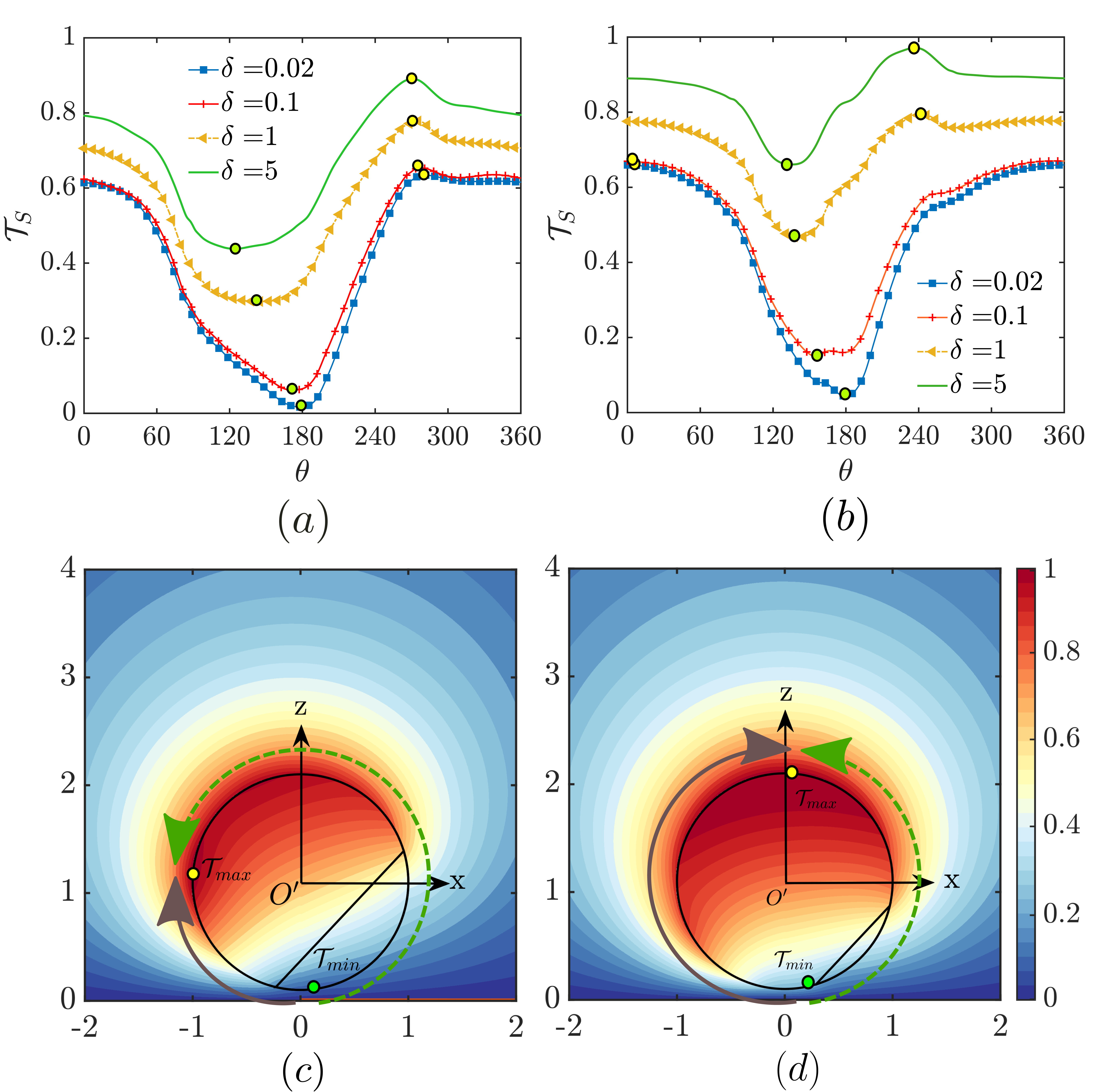}
    \vspace{+2ex}
    \caption[Swimmer surface temperature variation $ (\mathcal{T}_S) $ for different distances from the wall $ (\delta) $.]
    {Swimmer surface temperature variation $ (\mathcal{T}_S) $ for different distances from the wall $ (\delta) $. Subfigures (a) and (b) correspond to coverage angles of $ \varphi_{cap}=120^\circ$ and $ 150^\circ$, respectively. Subfigures (c), (d): The solid-black and green-dashed arrows indicate the direction of surface flow. We have chosen $ \varphi_{cap}=120^\circ$ and $ 150^\circ$, respectively in the two figures. The background colours denote the scaled temperature field.  The other parameters are $ \mathcal{K}=1,\,\delta=0.1$ and $ \theta_p=135^\circ$. The green and yellow bubbles denote the minimum and maximum temperature locations on the plots, respectively.}
    \label{fig:tcap_var_tmax_loc}
\end{figure}
\begin{figure}[t!]
    \centering
    \begin{subfigure}{0.43\textwidth}
        \centering
        \includegraphics[width=1.1\textwidth]{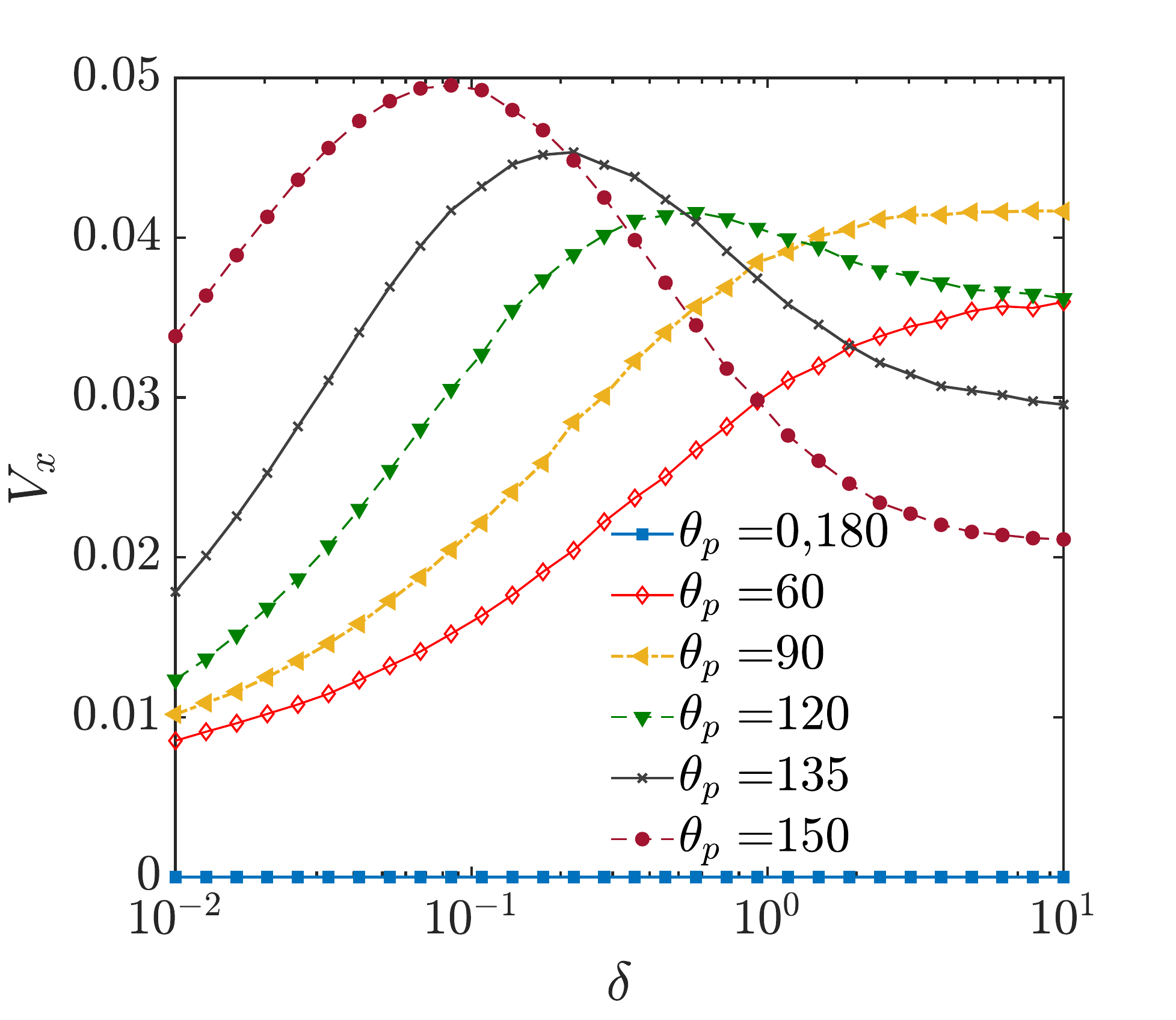}
                \vspace{5ex}
        \caption{} 
        \label{fig:Vx_vs_delta_vary_tp_tcap_150_K_1}
    \end{subfigure}    
    \begin{subfigure}{0.43\textwidth}
        \centering
        \includegraphics[width=1.1\textwidth]{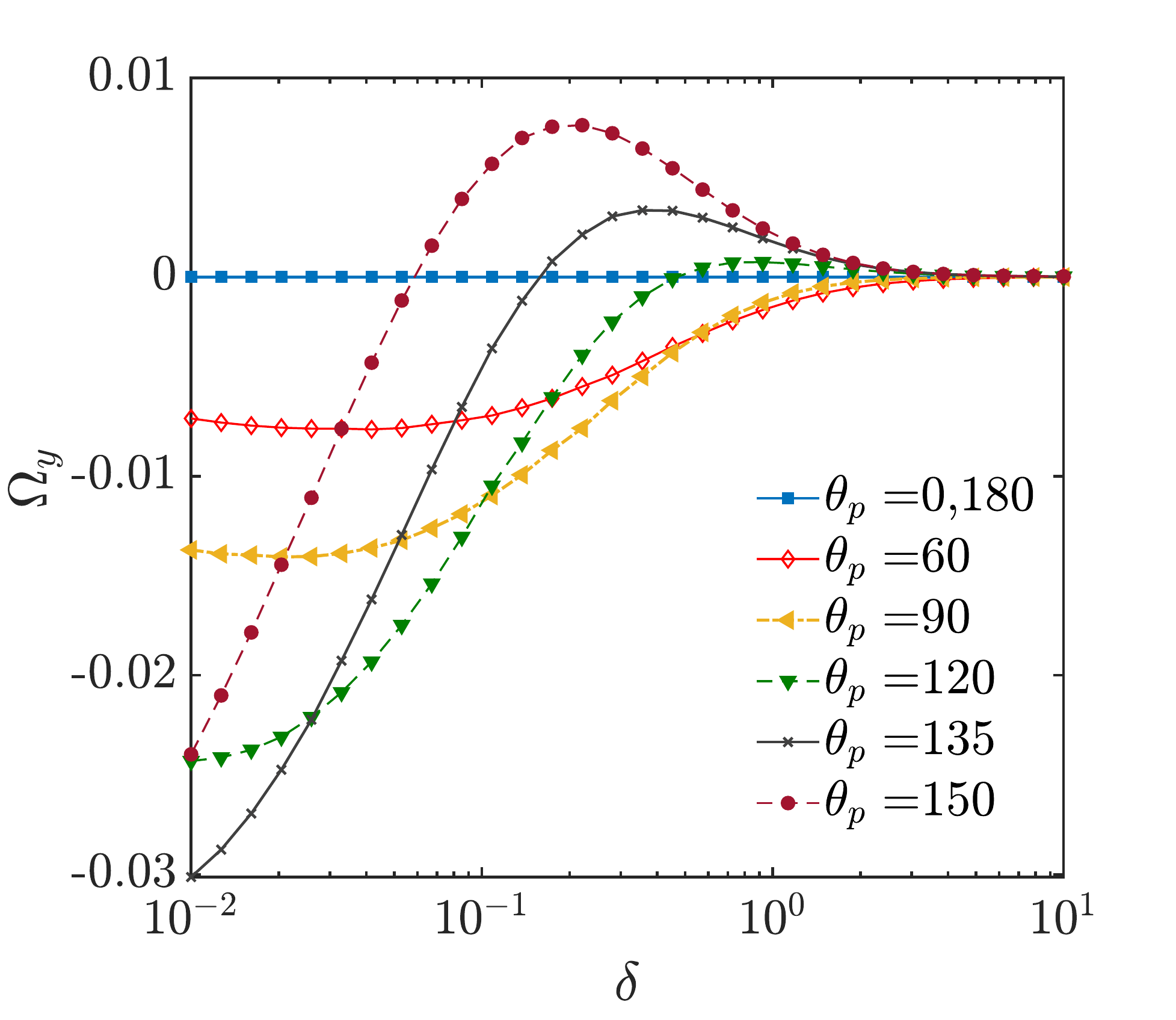}
                \vspace{5ex}
        \caption{} 
        \label{fig:Wy_vs_delta_vary_tp_tcap_150_K_1}
    \end{subfigure}
    \caption[Variation of wall parallel translational and rotational velocities of the microswimmer with distance from the wall $ (\delta) $ for different orientation angles $ (\theta_{p})$.]
    {Variation of wall parallel translational and rotational velocities of the microswimmer with distance from the wall $ (\delta) $ for different orientation angles $ (\theta_{p})$. Other parameters are chosen as $ \varphi_\text{cap} =150 ^{\circ}$ and $ \mathcal{K} =1$.} 
    \vspace*{6in}
    \label{fig:Vx_Wy_tcap_150_K_1}
\end{figure}

\end{appendices}
\FloatBarrier
\bibliographystyle{jfm2}
\begingroup
\def\bibfont{\small}
\setlength{\bibsep}{1.25ex}

\endgroup
\end{document}